\documentclass[acmsmall]{acmart} 

\usepackage{appendix}
\usepackage{fancybox}
\usepackage{enumitem}
\usepackage{graphicx}    
\usepackage{multirow}
\usepackage{subcaption}
\usepackage[utf8]{inputenc}

\definecolor{graybg}{RGB}{245, 245, 245} 
\definecolor{blackframe}{RGB}{0, 0, 0}   

\AtBeginDocument{%
  }

\setcopyright{cc}
\setcctype{by}
\acmJournal{PACMHCI}
\acmYear{2026} \acmVolume{10} \acmNumber{5} \acmArticle{MHCI3122}
\acmMonth{8} \acmDOI{10.1145/3821661}

\author{Pragya Singh}
\orcid{0000-0003-3933-2224}
\affiliation{%
  \institution{IIIT-Delhi}
  \city{New Delhi}
  \country{India}}
\email{pragyas@iiitd.ac.in}

\author{Prashasti Gupta}
\authornote{These authors contributed equally to this work.}
\orcid{0009-0002-2627-9145}
\affiliation{%
  \institution{IIIT-Delhi}
  \city{New Delhi}
  \country{India}}
\email{prashasti21346@iiitd.ac.in}

\author{Hitesh Bhandari}
\authornotemark[1]
\orcid{0009-0000-6782-1856}
\affiliation{%
  \institution{IIIT-Delhi}
  \city{New Delhi}
  \country{India}}
\email{hitesh20436@iiitd.ac.in}

\author{Kanishk Goel}
\orcid{}
\affiliation{%
  \institution{IIIT-Delhi}
  \city{New Delhi}
  \country{India}}
\email{kanishk21325@iiitd.ac.in}

\author{Mohan Kumar}
\orcid{0000-0002-0286-6997}
\affiliation{%
\institution{RIT}
  \city{Rochester}
  \country{New York, US}}
\email{mjkvcs@rit.edu}

\author{Pushpendra Singh}
\orcid{0000-0003-2152-1027}
\affiliation{%
\institution{IIIT-Delhi}
  \city{New Delhi}
  \country{India}}
\email{psingh@iiitd.ac.in}

\begin{document}

\title{Scales, Reflections, and Conversations: A Multi-Modal Approach to Emotion Annotation}


\begin{abstract}

Mental health concerns are increasing worldwide, highlighting the need for interventions that support everyday emotional well-being. Prior work has demonstrated the potential of wearable and mobile technologies to deliver data-driven interventions. However, developing effective data-driven systems requires access to emotion data that captures individuals’ emotional variability and change in everyday contexts. Existing approaches to data collection largely rely on frequent, prescheduled prompts and predefined scales or questionnaires. These methods often fail to account for participants’ availability, agency, or the complexity of their emotional experiences, resulting in shallow, context-poor data. In this paper, we present a feasibility study of a participant-centric, multimodal emotion-annotation application designed around users’ emotional intensity and availability. Our findings show how multimodal emotion logging can shape participants’ experiences and data logging behaviors, and demonstrate its potential to support the collection of richer, more nuanced emotion data.

\end{abstract}





\maketitle

\section{Introduction}

Emotions play a central role in shaping human experience, influencing everyday behavior, decision-making, quality of daily life, and overall mental well-being. Consequently, understanding and supporting how people experience, regulate, and reflect on their emotions through technology-enabled interventions has become an important area of research within human–computer interaction (HCI)  \cite{10.1145/3569898, 10.1145/3491102.3517573, wang2014studentlife}. This growing interest is driven by the widespread adoption of mobile devices, the increasing availability of affordable, sensor-equipped wearables, and advances in data-driven methodologies. 
Designing effective data-driven systems for everyday contexts hinges on their ability to accurately interpret and differentiate among diverse biomarkers and digital proxies as emotional states fluctuate. However, aligning variations in behavioral signals with underlying emotional changes remains inherently difficult, given the subjective nature of emotions and the indirect relationship between observable proxies and lived experiences. Consequently, researchers often rely on self-reports as the ground truth for labeling proxy-based emotion data \cite{saganowski_emotion_2023}. This reliance underscores the importance of carefully designing self-reporting systems that enable users to label their emotions in situ.

Despite their importance, collecting ecologically valid and nuanced emotional self-reports remains a persistent challenge. Such systems must balance being lightweight and unobtrusive with the need to capture sufficiently rich and high-quality data to support meaningful analysis, learning, and intervention. At the same time, they must mitigate common sources of labeling bias, including recall bias, social desirability bias, and negativity bias.
Prior work highlights several barriers to achieving this balance between data requirements and user needs  \cite{9434231}. First, frequent self-reporting can disrupt emotional privacy practices, raising concerns about how and when sensitive data is shared \cite{10.1145/3544548.3580950, 10.1145/3579600}. Second, differences in individuals’ ability to identify and articulate emotions introduce variability in how emotions are interpreted and labeled, contributing to label misalignment with emotional states \cite{10.1145/3749519, 10.1145/3711093}. Third, the act of self-reporting itself can shape or interfere with emotional experiences, leading to reactivity effects \cite{kelley_self-tracking_2017, kang2022understanding}. Finally, sustaining user engagement and motivation remains difficult, particularly when reporting mechanisms introduce friction into everyday routines \cite{10.1145/3025453.3025750, 10.1145/3749541, 10.1145/3191735}.
Consequently, researchers have explored a range of complementary methods to address these challenges in a combined effort to improve emotion data collection. These approaches include ecological momentary assessments \cite{10.1145/3706598.3714086, busso2025diversityone, 10.1145/3749541, yau2022tiles}, experience sampling \cite{10.1145/3123988}, day reconstruction methods \cite{10.1038/s41597-021-00945-4, busso2025diversityone}, digital phenotyping \cite{10.1038/s41746-018-0074-9}, digital diaries \cite{10.1145/3675094.3677550, 10.1145/2858036.2858360}, and journaling approaches \cite{10.1145/3613904.3642937, 10.1145/3699761}, often combining multiple techniques to balance ecological validity, user burden, and data fidelity.

However, much of the prior work on EMA approaches for capturing emotion in everyday contexts continues to rely on a single expression modality, most commonly rating scales, structured questionnaires, or, more recently, LLM-based scaffolding \cite{10.1145/3749519, 10.1145/3706598.3713732, busso2025diversityone, 10.1145/3610892, wang2014studentlife, 10.1145/3613904.3642662}. Across these approaches, a common assumption is that a single interaction format can sufficiently capture emotional experience in everyday life. However, research has shown that emotional expression is not uniform \cite{sias2016emotions, sacharin2012perception, fischer2004gender}, and it varies across situations and individuals. And how people express emotions is inherently subjective and influenced by a range of contextual factors and personal characteristics. As a result, individuals may not always be able to adequately express their emotions through a single interaction modality. Recent works have also highlighted that scale-based or questionnaire methods are often insufficient for conveying the full range and nuance of emotional experiences \cite{10.1145/3123988, 10.1145/3749519, 10.1145/3711093, singh2024eevr}. For example, complex emotional states, such as sadness intertwined with or masked by anger and frustration, often require more expressive space than simpler visible cue-based emotions like happiness. Additionally, both participants and mental health professionals have observed that individuals often lack the exact vocabulary to describe complex, overlapping, or abstract emotional states, which limits the effectiveness of predefined scales \cite{10.1145/3749519}.
Beyond limiting opportunities for emotional expression, single-modality systems can also constrain datasets and have downstream implications for AI models trained on them \cite{10.1145/3711093}. In particular, labels can fail to capture participants’ true emotional states, creating data misalignment \cite{singhfeel}. As a result, models trained on these datasets may not generalize well to real-world settings, where emotional expression is more heterogeneous and context-dependent \cite{singh2024eevr}. This limitation highlights the need for more adaptable annotation mechanisms, especially given that prior work has already shown improvements in model performance from richer self-reports and contextual information \cite{zhang2025sensorlm, singh2024eevr, NEURIPS2023_5f09bfe6}.

Overall, our prior discussion points highlight the need for more flexible participant-centric, multimodal approaches to collect emotional self-reports that can balance user burden with opportunities for expression while remaining scalable in real-world deployments  \cite{9434231}. Motivated by this, we designed a multimodal emotion self-reporting prototype system that supports flexibility in how users express their emotions. Furthermore, to explore how such flexibility shapes user behavior and emotion-reporting data, this paper presents a one-week feasibility study using our EMA prototype (see section \ref{application_design} for more details). We deployed our application in the field for seven days with 33 participants.
The study is guided by the following research questions:

\begin{enumerate}[label=\textbf{RQ\arabic*.}]
\item How do users engage with multimodal emotion logging in everyday contexts?
\item How does multimodal logging support varying levels of expressive elaboration, emotional complexity, and contextual grounding in emotion self-reports?
\end{enumerate}
Our formative in-field deployment investigates how users engage with an EMA system that supports flexible prompting schedules and multiple modalities for in-situ emotion logging. Furthermore, our findings demonstrate that supporting modality switching is not merely a usability enhancement but a mechanism that enables users to adapt expression to situational constraints and cognitive load. Overall, our results provide empirical evidence that multimodal and user-adaptive emotion logging systems are better suited to capturing the heterogeneity and situated nature of everyday emotional experiences than single-modality approaches often used in existing EMA approaches. We translate these findings into design implications for HCI systems that aim to support flexible, low-friction, and expressive forms of emotion self-report. Finally, we position our system as a design probe for future HCI research on emotion-aware interfaces. By foregrounding adaptability and agency, our work lays the foundation for developing more naturalistic, user-aligned tools for emotion self-reporting in everyday settings.

\section{Related Work}

\subsection{History of Emotion Self-Reporting} 

Researchers have long drawn on "psychological theories of emotion" to guide self-reporting scales and questionnaire designs. Categorical approaches, such as Ekman’s theory of basic emotions \cite{ekman1992there} and Plutchik’s model \cite{plutchik1982psychoevolutionary}, have informed the development of basic emotion scales that provide participants with predefined sets of discrete emotion labels to articulate their emotions. Complementing this perspective, dimensional approaches emphasize underlying affective dimensions for capturing emotional experiences. Famously, the Positive and Negative Affect Schedule (PANAS) \cite{watson1994panas} captures affect along positive and negative valence, while the Self-Assessment Manikin (SAM) \cite{bradley1994measuring} offers a non-verbal representation of valence, arousal, and dominance. More recently, appraisal theories \cite{roseman2001appraisal} have highlighted that individuals evaluate and interpret emotional events in context, thereby guiding the development of appraisal-based approaches \cite{larradet2019appraisal} for collecting emotional self-reports. Collectively, these theories have shaped the designs of available in-situ emotion self-reporting tools. 
Ecological Momentary Assessments (EMAs), also referred to as Experience Sampling Methods, have been a foundational approach for collecting emotion data in naturalistic settings that go beyond lab-based self-reporting methods \cite{10.1145/3191735}. These EMAs are often either interval-contingent, where a periodic time interval is set for prompting; signal-contingent, where the researcher decides the prompting schedule for emotion self-reports; event-contingent, where prompting is based on an event \cite{wheeler1991self}; or a hybrid combining these approaches. Prior research often employs these EMA approaches to collect self-reports of emotions \cite{wang2014studentlife, xu_globem_2022, 10.1145/3712289}.
A range of EMA toolkits has been designed in the past, including MindLamp \cite{vaidyam2022enabling}, AWARE \cite{ferreira2015aware}, AWARE-Light \cite{van2023aware}, PACO \cite{paco}, SensingKit \cite{katevas2016sensingkit}, mEMA \cite{ilumivu_mema}, ExperienceSampler \cite{thai2018experiencesampler}, and MobileQ \cite{meers2020mobileq}. 
However, most of these toolkits still face several challenges that limit their effectiveness. First, many toolkits capture emotional labels without adequate contextual information, making it difficult to interpret the context behind those changes in emotional states, which limits the ability to build robust predictive models \cite{10.1145/3191735}. Second, many EMA tools lack contextual awareness, leading to participant fatigue and reduced motivation \cite{10.1145/3706598.3714086, stone2023evaluation}. Third, these systems often offer limited opportunities for participants to express their emotions in situ \cite{10.1145/3749519}.
As a result, they lack usability or relevance to users’ lived experiences, reducing engagement and compromising the reliability of self-reports \cite{10.1145/3706598.3713732, stone2023evaluation}. These issues underscore the need for more participant-centered approaches to emotion self-reporting.

\begin{table*}[ht]
\centering
\begin{tabular}{p{3.5cm}p{4.5cm}p{4.5cm}}
\hline
\textbf{Dataset} & \textbf{Collection Setting} & \textbf{Self-Reporting Approach} \\
\hline
\textbf{NURSE} \cite{10.1038/s41597-022-01361-y} & Healthcare workers during COVID-19 & Custom Stress Questionnaire \\
\textbf{G-REx} \cite{bota2024real} & Long movie viewing sessions & Post-Hoc SAM Scale Based Tool \\
\textbf{Laureate} \cite{10.1145/3610892} & University setting with student academic routines & Custom EMA (PANAVA-KS, physical activity, breakfast ingestion, caffeine intake, study-time and sleep quality) \\
\textbf{StudentLife} \cite{wang2014studentlife} & University campus life over multiple weeks & Photographic Affect Meter (PAM) EMA, Single-item Stress EMA \\
\textbf{GLOBEM} \cite{xu_globem_2022} & Naturalistic daily experiences across diverse locations & EMA Survey (PHQ-4, PSS-4, PANAS), and Pre-Post Survey \\
\textbf{TILES} \cite{yau2022tiles, yau2022tiles} & Workplace monitoring in hospital environment & Single-item Stress EMA, Survey on daily stressors, work behaviors, and sleep \\
\textbf{SWEET Study} \cite{10.1038/s41746-018-0074-9} & Office workers’ daily routines in real-life settings &  EMA (Stress, Activity, Food and Beverage Consumption, Sleep Quality, and Gastro-intestinal Symptoms) \\
\textbf{DAPPER} \cite{10.1038/s41597-021-00945-4} & Daily life across varied settings & 20-Item ESM (Information about daily events, Participants' openness to sharing emotion, TIPI-C, PANAS), DRM with Open-ended Question \\
\textbf{K-EmoPhone} \cite{kang2023k} & Daily life across varied settings & Custom Questionnaire (Valence, Arousal, Attention, Stress, Emotion Duration, Task Disturbance, Emotion Change)\\
\textbf{Diversity One}  \cite{busso2025diversityone} & University students across eight countries over four weeks & Morning and Evening Diaries (sleep quality and daily expectations), Time Diaries, and Snacks Diaries\\
\textbf{LifeSnaps} \cite{yfantidou_lifesnaps_2022} & University Students from four european countries over 4 months & Step goal EMA, Context and Mood EMA\\
\hline
\end{tabular}
\caption{A Review of Emotion Datasets Collected in Naturalistic and Semi-Naturalistic Settings}
\label{tab:real_datasets_rephrased}
\end{table*}

\subsection{Participant-Centric Emotion Logging} 

To overcome the challenges posed by simplistic self-reporting toolkits, researchers have turned their attention to more interactive and user-centric methods \cite{10.1145/3123988, 8668435, rajcic2020mirror, wang2018mirroru, hook2009affective}. A variety of tools and systems have emerged, such as Reconexp \cite{10.1145/1409240.1409316}, which offers both mobile and web-based interfaces to facilitate emotion reporting, mirrorU \cite{10.1145/3170427.3188517}, which promotes reflective writing by prompting users with memory cues, and Find the Bot \cite{10.1145/3613904.3642880}, which utilizes a gamified web-platform. Further interventions include PResUP \cite{10.1145/3678569}, which encourages opportunistic emotion reporting throughout the day, and Mirror Ritual \cite{10.1145/3313831.3376625}, which combines facial emotion recognition with AI-generated poetry to stimulate emotional reflection. Mindnotes \cite{chanda_mindnotes_2021}, a mobile-based tool designed to support emotion articulation beyond stigma.
Other context-driven techniques include circadian rhythm-assisted methods \cite{stone2006population}, technology-assisted reconstruction (TAR) framework \cite{karapanos2012beyond}, which uses passively collected data to support end-of-day emotional reflection and self-reports, and Mirror Hearts \cite{10.1145/3544549.3585607} offers an AI-powered third-person perspective to enhance self-awareness during emotion reporting. Additionally, an interactive versions of structured scales are also designed. These include traditional scales such as the Affect Grid \cite{russell1989affect}, the Differential Emotions Scale \cite{boyle1984reliability}, the Premo \cite{desmet2003measuring}, and the Photographic Affect Meter (PAM) \cite{10.1145/1978942.1979047}, in which users indicate their emotional states by selecting representative images or blocks. More recently, a mobile-friendly version of the Geneva Emotion Wheel has also been introduced \cite{simonazzi2021geneva}. 
Recently, large language models (LLMs) have also been explored for scaffolding emotional self-reflections and AI-journaling. Methods like DiaryHelper \cite{li2024diaryhelperexploringuseautomatic}, Mindshift \cite{10.1145/3613904.3642790}, Diarymate \cite{10.1145/3613904.3642693}, and Mindscape \cite{10.1145/3699761} explores in-context journaling and conversational interfaces for emotion reflecting. These developments suggest a significant evolution in how emotion self-reporting should also support participants' needs. There is a clear shift from rigid, scale-based methods toward more naturalistic, context-aware approaches that integrate seamlessly into daily life. Overall, emotion data collection is becoming more embedded, reflective, and empathetic, with the potential not only to generate more meaningful data but also to support emotional well-being. In this study, we aim to further investigate the potential of incorporating participant-centric features into emotion self-reporting tools to enable more authentic data collection in real-life settings. 

\subsection{Emotion Data Collection in-the-Wild}

Emotion data collection studies commonly collect multimodal data, including various physiological signals, behavioral data, and self-reports, to derive insights into users’ emotions across diverse contexts. Traditionally, the data was collected in lab environments using standardized protocols, including stimulus-driven elicitation (e.g., video clips, or psychological tasks) and structured self-assessment tools like the SAM, PANAS, or standardized mental-health scales (e.g., STAI, PHQ-9) \cite{tabbaa2021vreed, miranda-correa_amigos_2021, 10.1145/3242969.3242985, saganowski2022emognition}. Lab-based datasets have advanced our understanding of emotions but face key limitations for real-world use, such as a lack of ecological validity \cite{9779458, 10.1145/3711093} and the collection of only brief snapshots rather than capturing the temporal dynamics of emotions. Recently, research has shifted toward collecting data in the wild. Large-scale studies, such as StudentLife \cite{wang2014studentlife}, GLOBEM \cite{xu_globem_2022}, DiversityOne \cite{10.1145/3712289}, and LAUREATE \cite{10.1145/3610892}, capture emotions and behaviors in real-world contexts over weeks or months, enabling longitudinal insights into behavioral patterns. Despite these advances, real-world emotion data collection faces persistent challenges such as participant burden and prompt fatigue, which lead to non-reliable or missing data \cite{10.1145/3706598.3714086, 10.1145/3749519, 10.1145/3491102.3501944}. Moreover, these studies suffer from participants' biases and errors and often lack contextual details \cite{das2022semantic, 10.1145/3711093}, which leads to poor-performing emotion recognition models \cite{singhfeel}. These limitations highlight the need to design more contextually rich datasets that capture emotional nuances to develop models beyond lab settings \cite{singh2024eevr, singhfeel}. 

\begin{figure}[ht]
    \centering
    \includegraphics[width=\columnwidth]{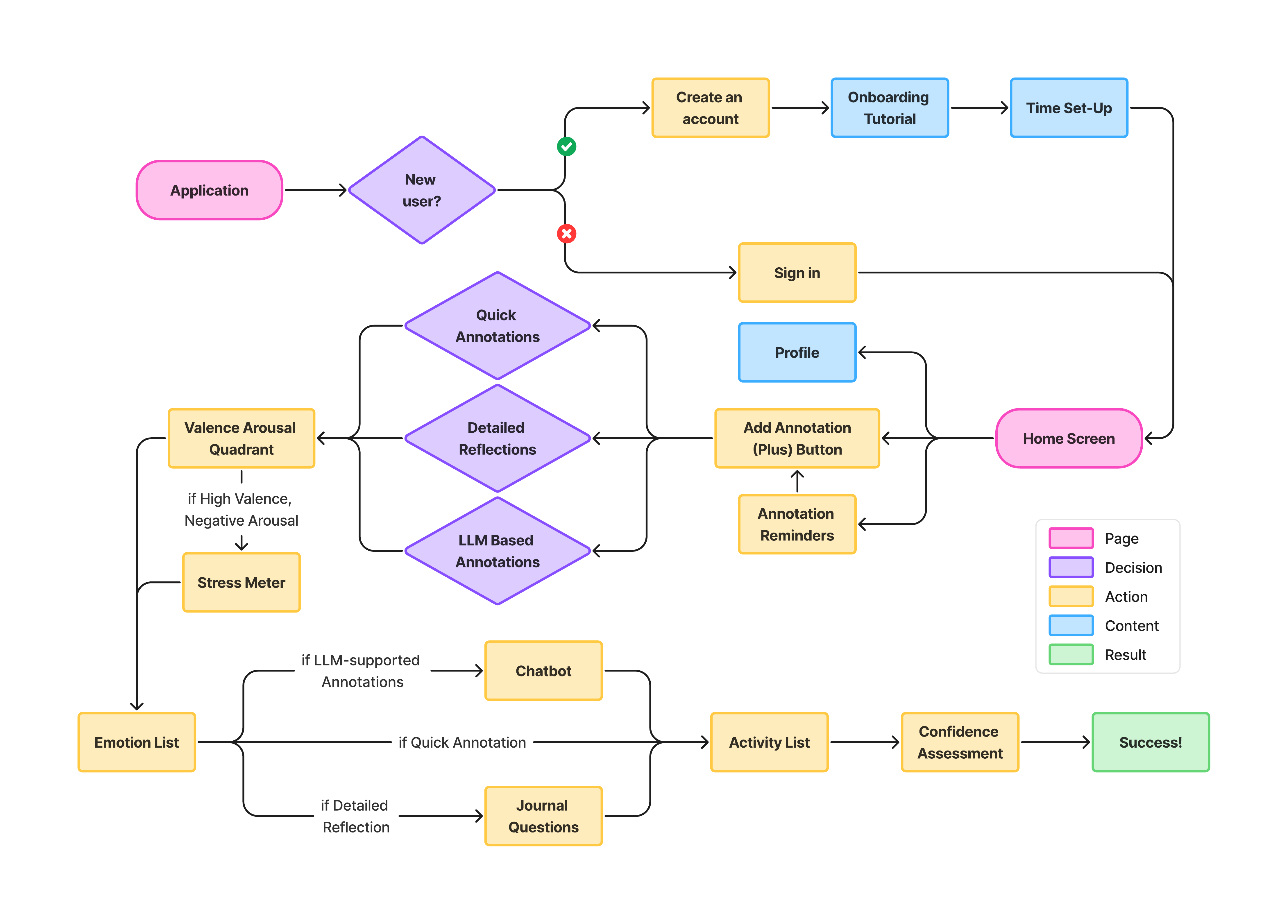} 
    \caption{An Illustration of our Application Flow (Best viewed in color).}
    \Description{A diagram showing the flow of the application, including main components and their interactions.}
    \label{fig:appflow}
\end{figure}

\section{Application Design: Overview}\label{application_design}

Formative user-centered research \cite{10.1007/978-3-030-78462-1_10} and behavioral theories such as self-determination theory \cite{ryan2000self} consistently highlight that intrinsic motivation is strongly linked to perceived autonomy and agency. In data collection contexts, providing users with greater control over when and how they contribute data has been shown to improve both engagement and willingness to share information \cite{chang2017investigation, srinivas2019context}. In addition, prior work emphasizes the importance of adaptive and personalized designs to better accommodate diverse user needs and contexts \cite{wang2024designing}. Guided by these principles, we developed our prototype. We took design insights from prior user-centered emotion logging systems \cite{10.1145/2750858.2807524, 10.1145/3749519, 10.1145/3447526.3472039, 10.1145/3711093, 10.1145/3123988, li2024diaryhelperexploringuseautomatic}, as well as commercial applications such as Apple Health, Daylio, and Moodflow, for developing our prototype. 
Our prototype integrates the following set of design features aimed at supporting flexible and multi-modal emotion reporting:

\begin{enumerate}
\item \textbf{User-configurable prompting and impromptu logging}: The system supports both user-defined reminders and on-demand logging, enabling individuals to record emotions at self-selected times as well as in-the-moment self-reporting. This design choice supports flexibility, accommodates varying daily routines, and reduces reliance on externally imposed schedules.

\item \textbf{Multi-modal self-reporting options}: To accommodate diverse expressive needs, the system offers multiple reporting modalities. This allows users to select the mode that best aligns with their context, cognitive load, and preferred level of expression, while also accounting for variability in emotional vocabulary and articulation.

\item \textbf{Supporting contextual reporting}: The system incorporates multiple supportive mechanisms to enable richer contextual annotation.
\end{enumerate}

Our application design is illustrated in Figure \ref{fig:appflow}. More details on the technical implementation are provided in Appendix \ref{tech}. Next, we present the system design along with the underlying design rationales.

\begin{figure}[t]
    \centering
    \includegraphics[width=\columnwidth,keepaspectratio]{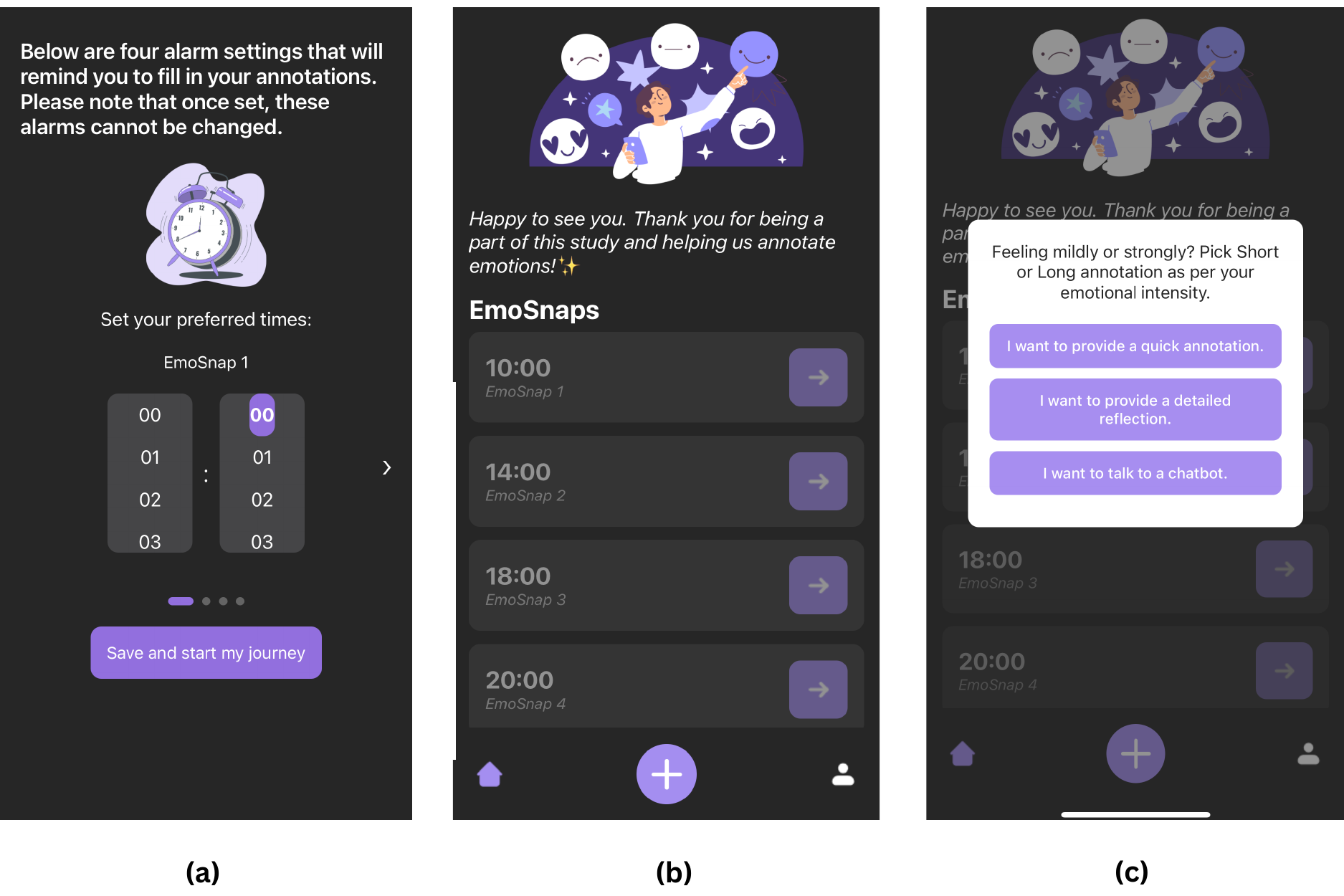}
    \caption{(a) Time-slot selection screen where users choose four daily notification reminders. (b) Home screen displaying the selected times and a floating action button for on-demand emotion logging. (c) Modal window shown when the button is tapped or a notification is opened, offering three emotion self-report options (Best viewed in color).}
    \Description{The figure illustrates the onboarding and interaction flow of the emotion logging application. (a) The time-slot selection screen allows users to configure four daily reminder notifications by choosing preferred times throughout the day. (b) The home screen displays the selected reminder schedule and includes a floating action button that enables users to initiate an emotion log at any time. (c) When a reminder notification is opened or the floating action button is tapped, a modal window appears presenting three available emotion self-report modalities from which users can choose to record their emotional experiences.}
    \label{fig:annotation}
\end{figure}

\subsection{Onboarding Module} 
The onboarding module consisted of: (1) \textbf{Interactive Tutorial}, which introduced the concept of emotion annotation and the scales used in our application, including arousal–valence dimensions, the four-quadrant system derived from it, and the perceived-stress scale (see Figure \ref{fig:tutorial}). The tutorial was added to support participants' training and reduce interpretation biases \cite{stone2023evaluation}, and (2) \textbf{Time-slot Selection} (see Figure \ref{fig:annotation}a), where participants scheduled four daily slots aligned with their routines. Notifications were delivered at these times, integrating self-reporting into daily life. Following prior works \cite{kang_k-emophone_2023, shui_dataset_2021}, we limited reminders to four per day to balance data requirements with participant burden.

\subsection{Home Screen and Modality Selection}

After completing the tutorial and selecting their slots, participants arrived at the home screen, which displayed four prescheduled slots represented as “Emosnaps” (see Figure \ref{fig:annotation}b). Each Emosnap was locked by default and unlocked one hour before the scheduled time, remaining available for one hour afterward to allow flexible yet temporally relevant reporting. Once the window closed, the Emosnap was locked again to preserve temporal fidelity. We anchored prompts to user-defined routines to leverage established behavioral patterns, supporting habit formation through temporal consistency while preserving user autonomy in scheduling. The home screen also featured a floating action button for spontaneous entries, enabling participants to log their emotions whenever they felt something significant happened. This dual structure, combining scheduled prompts with on-demand logging, was therefore intended to support both structured recall and event-driven reporting, thereby aiming to increase ecological coverage of emotional experiences while preserving user autonomy and reducing interaction friction.

\subsection{Self-reporting Methods}

To study our research questions, in our prototype, we included three self-reporting approaches (see figure \ref{fig:annotation}c): Quick Mode, Detailed Reflections, and LLM-supported Annotations. Each of these approaches was designed to provide users with multiple modes to express their emotions, depending on their context, cognitive load, and expression needs.

\subsubsection{\textbf{Quick Mode}} 
The inclusion of this scale-based mode was motivated by its widespread use in traditional EMA systems as a lightweight and standardized mechanism for emotion self-reports. However, to address its known limitations, we extended it with a multi-select emotion list alongside the arousal–valence scale. This design allows users to report multiple concurrent emotional descriptors, thereby better capturing mixed, overlapping, and co-occurring affective states that are difficult to represent within a purely dimensional framework. In addition, the multi-select context list and confidence rating were incorporated to enrich each report with contextual and subjective uncertainty information. Together, these elements were intended to increase the interpretability of self-reports by providing additional information for downstream modeling and analysis. It takes approximately one minute to self-report in this mode. The flow of this annotation method is as follows:
\begin{enumerate}
    \item \textbf{Quadrant Selection}: Users start with the arousal–valence quadrant screen (based on Russell’s Circumplex Model of Affect \cite{russell1980circumplex}. This color-coded screen has four quadrants, each representing a combination of arousal and valence. For example, the red quadrant (high arousal–negative valence) reflects emotions like anxiety or anger, while the yellow quadrant (high arousal–positive valence) represents excitement or happiness. The design was informed by prior emotion-assessment tools, like the SAM \cite{bradley1994measuring}, Affect Grid \cite{russell_affect_1989}, Geneva Emotion Wheel \cite{simonazzi2021geneva}, and the Photographic Affect Meter \cite{10.1145/1978942.1979047}.
    \item \textbf{Stress Scale} (conditional): After the quadrant screen, users were shown the perceived stress scale (PSS) only if they selected the high arousal–negative valence quadrant, which corresponds to stress-related emotions. This scale, adapted from the widely used 10-item PSS \cite{reis2010perceived}, was included to measure stress intensity, often collected separately in emotion datasets \cite{10.1145/3242969.3242985}. To improve usability, each numerical value was paired with a short descriptive phrase, for example, seven represents moderately high stress.
    \item \textbf{Emotion List}: Next, users were shown a curated list of emotions corresponding to their selected quadrant (see Table \ref{tab:valence-arousal-emotions}). They could select multiple labels to describe their states (see figure \ref{fig:annotation2}a). This screen was designed to provide both a guided vocabulary and the flexibility to capture concurrent or overlapping emotions alongside quadrant labels \cite{10.1145/3749519}.
    \item \textbf{Contextual Factors}: After selecting their emotions, users chose from a predefined list of activities they had engaged in since their last log (see Table \ref{tab:contextual-factors}). The list, covering domains such as health, physical activity, medication, and environmental influences \cite{10.1038/s41746-018-0074-9}, was designed to help users reflect on possible triggers of their emotions.
    \item \textbf{Confidence Rating}: Finally, users can rate their confidence in the accuracy of their annotation on a 5-point scale (see Figure \ref{fig:annotation2}d). This step was designed to encourage self-reflection \cite{schroder2006first} while also providing researchers with an additional indicator of data reliability \cite{10.1145/3749519}.
\end{enumerate}

\subsubsection{\textbf{Detailed Reflections}} 
The detailed mode was designed to capture more complex and context-rich emotional experiences that cannot be adequately expressed through quick mode (see Figure \ref{fig:annotation2}b). This design draws on principles of reflective practice in HCI, where prompts can facilitate deeper sense-making while maintaining consistency across entries. These prompts can help users articulate their thoughts, situational triggers, and emotional interpretations without facing the cognitive burden of open-ended reflection. After completing the initial scale-based categorization (quadrant selection, stress scale (if applicable), and emotion list), users were presented with four open-ended journaling prompts, delivered across separate screens. The prompts were as follows:
    \begin{enumerate}
        \item How would you describe what you're feeling right now?
        \item Did your body give you any clues about this feeling?
        \item What do you think sparked this emotion?
        \item Can you pin down the moment or thought that started it?
    \end{enumerate}
The prompts were grounded in the ABC model of journaling \cite{malkinson2010cognitive}. To support flexibility, users could skip any question they preferred not to answer. To accommodate diverse expression styles, the detailed reflection mode also included multimedia options, allowing users to record audio or upload images \cite{10.1145/3706598.3714086}. Finally, consistent with the quick mode, users could also log their current activity and rate their confidence in the reflection, ensuring coherence and comparability across data.

\subsubsection{\textbf{LLM-Supported Reflections}} 
\begin{figure}[ht]
    \centering
    \includegraphics[width=\columnwidth]{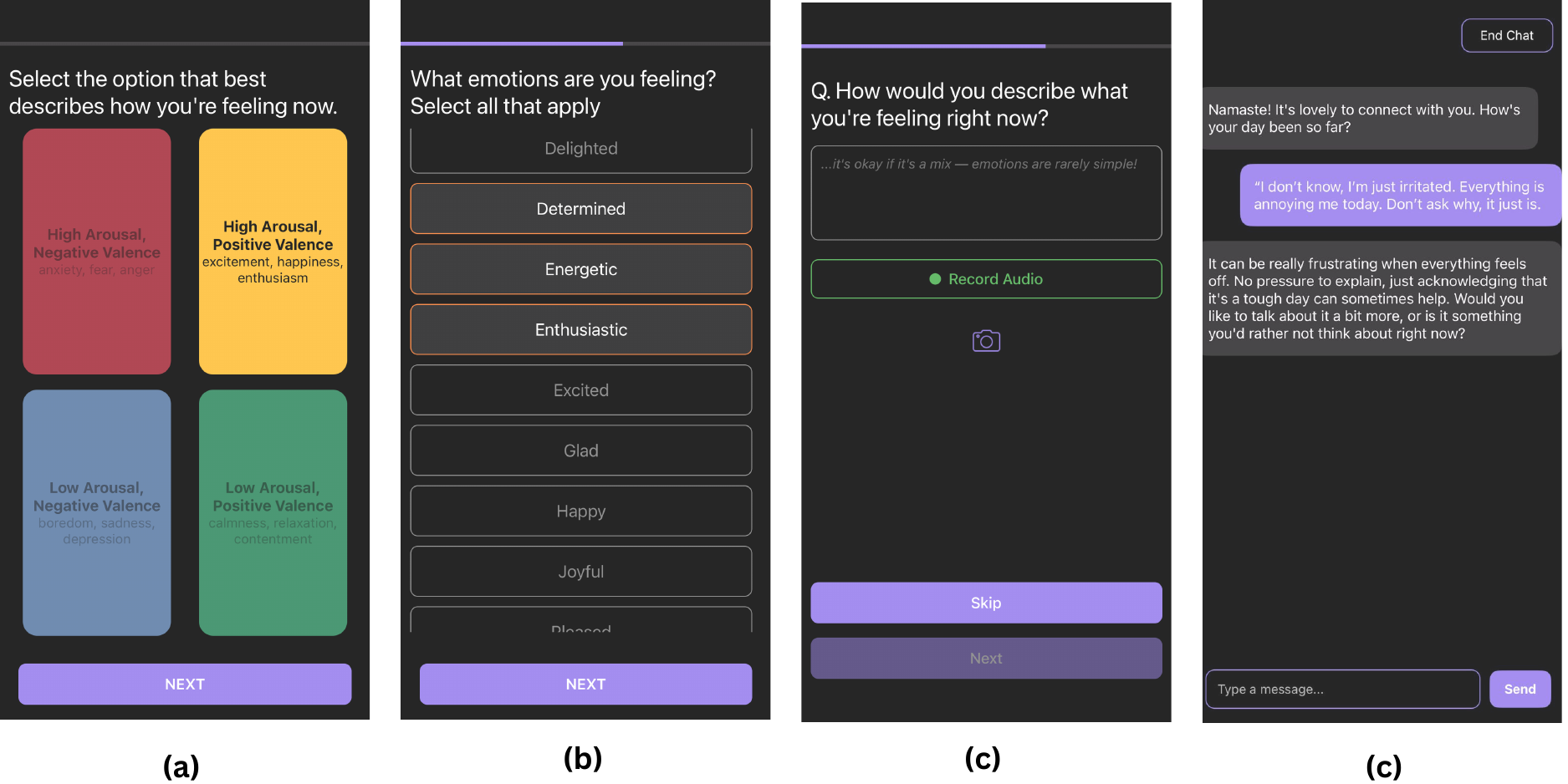}
    \caption{This figure displays the three reporting modes: (a–b) the quadrant screen and emotion list, (c) the detailed mode (Q1), and (d) the LLM-based mode (Best viewed in color).}
    \Description{The figure presents the three emotion reporting modalities available in the application. (a) The quadrant screen allows users to locate their emotional state within a two-dimensional emotion space. (b) The emotion list provides a set of emotion labels that users can select to further specify their feelings. (c) The detailed reporting mode presents a structured reflection prompt (Q1) that encourages users to elaborate on their emotional experiences. (d) The LLM-based conversational mode enables users to engage in a dialogue-driven interaction to describe and reflect on their emotions in a more open-ended manner.}
    \label{fig:annotation2}
\end{figure}

As a third self-reporting modality, the system includes a conversational interface that enables users to engage in dialogue while reporting emotions (see Figure \ref{fig:annotation2}c). The rationale for including this mode is to complement both quick-entry and structured journaling approaches by introducing interactive scaffolding for cases where emotions are ambiguous, evolving, or difficult to articulate. We hypothesize that, unlike one-way reporting, a conversational format can enable iterative clarification through prompts and follow-up questions, helping users progressively refine and externalize their emotional experiences. Given that most prior EMA systems rely on single-modality inputs, this mode is positioned as an exploratory extension to examine how the LLM-mediated annotation mode might support existing predefined format-based approaches for data collection. This design is further motivated by the recent adoption of LLMs in reflective journaling contexts, where they have shown promise in facilitating emotion articulation \cite{nepal2024contextual, 10.1145/3699761, 10.1145/3613904.3642693, 10.1145/3613904.3642937}. To maintain design consistency, the chatbot option followed the same sequence as the other methods: quadrant selection, conditional stress scale, emotion list, chatbot interaction, and activity selection and confidence rating. 

\begin{table}[t]
\centering
\begin{tabular}{ll}
\toprule
\textbf{Contextual Factor} & \textbf{Examples or Description} \\
\midrule
Physical Activity & Performed some physical activity \\
Temperature Change & Change in temperature (e.g., AC to outdoors) \\
Medication & Took some form of medication \\
Food Intake & Had food recently \\
Caffeine & Consumed caffeinated drinks \\
Alcohol/Sugar & Consumed alcohol or sugary drinks \\
Environment & Noisy, crowded, or chaotic surroundings \\
Health & Feeling unwell or in pain \\
Supplements & Took vitamins or supplements \\
Recreational Substances & Used substances like nicotine \\
Menstruation & Menstruating (if applicable) \\
None of the Above & No relevant contextual factor \\
\bottomrule
\end{tabular}
\caption{Contextual Factors List}
\label{tab:contextual-factors}
\end{table}

\begin{table}[]
\centering
\resizebox{\columnwidth}{!}{
\begin{tabular}{cccccccc}
\hline
\textbf{PID} & \multicolumn{1}{l}{\textbf{Age}} & \textbf{Gender} & \textbf{Education} & \textbf{Occupation} & \textbf{\begin{tabular}[c]{@{}c@{}}Mental Health \\ Diagnosis\end{tabular}} & \textbf{In Therapy} & \textbf{\begin{tabular}[c]{@{}c@{}}Emotional \\ Event\end{tabular}} \\ \hline
P1 & 26 & Male & Master's Degree & Student & No & No & Yes \\
P2 & 27 & Male & Bachelor's Degree & Phd Student & No & No & Yes \\
P3 & 23 & Female & Master's Degree & Phd Student & No & No & Yes \\
P4 & 23 & Male & Bachelor's Degree & Founder & No & No & Yes \\
P5 & 29 & Female & Bachelor's Degree & Software Engineer & No & No & Yes \\
P6 & 29 & Female & Master's Degree & Student & \begin{tabular}[c]{@{}c@{}}Schizophrenia, \\ Depression \\ and Anxiety\end{tabular} & Yes & Yes \\
P7 & 22 & Male & High school & Student & \begin{tabular}[c]{@{}c@{}}Acute Clinical \\ Depression and \\ Anxiety\end{tabular} & Yes & Yes \\
P8 & 23 & Female & Bachelor's Degree & Research Associate & Anxiety & Yes & No \\
P9 & 25 & Female & Master's Degree & Student & Anxiety & Yes & Yes \\
P10 & 28 & Male & Master's Degree & PhD Student & No & Yes & No \\
P11 & 23 & Male & Bachelor's Degree & Software Engineer & PTSD & Yes & Yes \\
P12 & 21 & Female & High school & Student & No & No & No \\
P13 & 29 & Male & Bachelor's Degree & Software Developer & No & No & No \\
P14 & 26 & Female & Master's Degree & Home Maker & No & No & Yes \\
P15 & 24 & Male & Bachelor's Degree & Software Engineer & No & No & Yes \\
P16 & 29 & Female & Master's Degree & Phd Student & No & Yes & Yes \\
P17 & 30 & Female & Master's Degree & Phd Student & No & No & Yes \\
P18 & 22 & Male & Bachelor's Degree & Software Engineer & No & No & No \\
P19 & 27 & Female & Master's Degree & Phd Student & No & No & Yes \\
P20 & 27 & Female & Master's Degree & Phd Student & No & No & No \\
P21 & 27 & Male & Bachelor's Degree & Phd Student & No & No & No \\
P22 & 21 & Female & High school & Student & No & No & Yes \\
P23 & 21 & Female & High school & Student & No & Yes & Yes \\
P24 & 22 & Female & High school & Designer & ADHD & No & Yes \\
P25 & 34 & Male & Bachelor's Degree & Freelancer & No & No & No \\
P26 & 21 & Female & High school & Student & No & No & Yes \\
P27 & 26 & Female & Master's Degree & Phd Student & No & No & Yes \\
P28 & 25 & Male & Bachelor's Degree & Phd Student & No & No & No \\
P29 & 37 & Female & Master's Degree & Manager & No & No & Yes \\
P30 & 22 & Female & Bachelor's Degree & Research Associate & No & Yes & Yes \\
P31 & 20 & Male & High school & Student & No & Yes & No \\
P32 & 22 & Male & High school & Software Developer & No & No & Yes \\
P33 & 27 & Female & Master's Degree & Phd Student & No & Yes & Yes \\ \hline
\end{tabular}
}
\caption{Participant Demographics and Mental Health Background. Emotional Events refer to participants’ recent experiences with significant emotional events.}
\label{participants}
\end{table}

\begin{table}[htbp]
\centering
\resizebox{\columnwidth}{!}{
\begin{tabular}{p{5cm}p{9cm}}
\hline
\textbf{Question} & \textbf{Response Summary (N=33)} \\
\hline
Daily routine & Very structured: 3 (9.1\%), \textbf{Somewhat structured: 22 (66.7\%)}, \\
& Unstructured: 8 (24.2\%) \\
Family dynamics & Supportive and emotionally open: 12 (36.4\%) \\
& \textbf{Supportive but not emotionally expressive: 14 (42.4\%)} \\
& Limited emotional support: 7 (21.2\%) \\
Work-life Balance & \textbf{Well: 14 (42.4\%)}, Moderately: 11 (33.3\%), Poorly: 8 (24.2\%) \\
\hline
Comfort Expressing Emotions  & Very comfortable: 1 (3.0\%), \textbf{Somewhat comfortable: 15 (45.5\%)}, \\
& Neutral: 10 (30.3\%), Somewhat uncomfortable: 7 (21.2\%), \\
& Very uncomfortable: 0 (0.0\%) \\
Concerned about others perception & Very concerned: 7 (21.2\%), \textbf{Somewhat concerned: 11 (33.3\%)}, Neutral: 10 (30.3\%), Not very concerned: 4 (12.1\%), Not concerned at all: 1 (3.0\%) \\
Emotions as a sign of weakness & Strongly agree: 2 (6.1\%), Somewhat agree: 3 (9.1\%), Neutral: 5 (15.2\%), \\
& Somewhat disagree: 9 (27.3\%), \textbf{Strongly disagree: 14 (42.4\%)} \\
\hline
Past Experience with Emotion Logging & \textbf{Yes}: 7, No: 26 \\
Alexithymia (TAS-20) & \textbf{Low}: 30 (90.9\%), High: 3 (9.1\%) \\
Cognitive Reappraisal (ERQ-6) &\textbf{High}: 27 (81.8\%), Low: 6 (18.2\%) \\
Expressive Suppression (ERQ-6) & \textbf{High}: 21 (63.6\%), Low: 12 (36.4\%) \\
Resilience (BRS-6) & \textbf{High}: 22 (66.7\%), Low: 11 (33.3\%) \\
\hline
\end{tabular}
}
\caption{Participant responses to daily support environment, emotional expression, and psychosocial measures.}
\label{tab:psychosocial_profile}
\end{table}

\section{Feasibility Study}


\textbf{Pre-Study Survey}: We administered a pre-study survey via email to all interested participants, which included informed consent and baseline questions on demographics, mental health history, prior counseling, recent emotional events, daily routines, family and work–life context, and comfort with emotional expression. These factors helped contextualize participants’ self-reporting behaviors, given known influences of routine, privacy, and social perception on EMA engagement \cite{10.1145/3749519, 10.1145/3191735, trampe2015emotions}. The survey also incorporated three standardized measures: the 20-item Toronto Alexithymia Scale (TAS-20) \cite{bagby2020twenty}, the 6-item Emotion Regulation Questionnaire (ERQ-6) \cite{preece2023emotion}, and the 6-item Brief Emotion Resilience Scale (ERS-6) \cite{smith2008brief}. These instruments captured individual differences in emotion identification, regulation, and resilience, which were important for interpreting how participants interacted with the emotion-logging system \cite{10.1145/3749519, 10.1145/3699761}. More details added in appendix \ref{prestudy}.


\textbf{Field Study Design}: The study ran for three weeks, with each participant using the application for one week based on availability. Participants were 20–37 years old (M = 25.42, SD = 3.98), including 14 men and 19 women. After completing the pre-study survey, all participants attended an \textbf{onboarding session} (in-person or online) covering installation, daily self-reporting procedures, and data privacy practices, reinforced later via email and a user manual \cite{10.1145/3749519}. During setup, participants selected four daily notification slots that aligned with their routines and were introduced to impromptu logging option via the floating-point button. Each participant used the app for a planned seven-day period, though some continued voluntarily for up to 11 days, citing its usefulness for tracking emotions. We analyzed all collected data to capture authentic engagement patterns. At study completion, participants completed a \textbf{feedback survey} assessing usability, relevance, and overall satisfaction \cite{10.1145/3699761} (More details added in appendix \ref{feedback_survey}), followed by an online \textbf{semi-structured exit interview} conducted via Zoom Pro (More details added in appendix \ref{interview questions}). The study received Institutional Review Board approval.

\textbf{Participants Recruitment}: We recruited participants using a mix of snowball sampling \cite{goodman1961snowball} and convenience sampling \cite{stratton2021population}, leveraging institutional emails and social media. Of the 50 individuals who expressed interest, 35 enrolled and 33 completed the one-week study, with two withdrawing on the first day due to scheduling conflicts. All participants were over 18, enrolled voluntarily without paid incentives, and provided informed consent. Participants were not incentivized, allowing us to observe engagement that more closely reflects authentic, voluntary use. No exclusion criteria related to mental health history or prior emotion-logging experience were applied. Our goal was to capture varied psychosocial profiles and emotional experiences. Table \ref{participants} summarizes demographics and mental health history, while Table \ref{tab:psychosocial_profile} details psychosocial profiles including support environments, emotional expression, alexithymia, regulation, and resilience.

\section{Analysis}
 
We adopted a mixed-methods approach combining descriptive quantitative analysis, mixed-effects modeling, and qualitative interpretation to examine how multimodal emotion logging shapes both user behavior and the expressive characteristics of self-reports in everyday contexts. Given the exploratory nature of this study, our goal was not to evaluate long-term compliance but to examine how specific design features (scheduling flexibility, multi-modality, multi-select emotion list, and media sharing) shape (1) user experiences, data-sharing behavior, and (2) the characteristics of the resulting emotion data. To address RQ1, focusing on user behavior, we investigate the following sets of exploratory questions:
\begin{itemize}
\item \textbf{E1}: How do impromptu and scheduled logging approaches differ in terms of supporting user flexibility and the differentiation of reported emotional experiences?
\item \textbf{E2}: How is annotation modality choice associated with temporal and affective context, reflecting users’ adaptation to situational constraints and cognitive load?
\item \textbf{E3}: How do individual characteristics (e.g., mental health history, emotional profiles, daily routines) relate to interaction patterns and modality preferences?
\end{itemize}

To examine E1–E3, we employed mixed-effects models with participant included as a random effect to account for repeated measures and inter-individual variability in logging behavior. We used this approach to appropriately model the nested structure of our data, where multiple observations are contributed by each participant across time and modalities. Fixed effects (e.g., scheduling type and annotation modality) and dependent variables were specified according to each exploratory question; further details are provided in the findings section. Furthermore we assessed model assumptions through visual inspection of residual-versus-fitted and Q–Q plots. These diagnostics did not indicate substantial violations of normality or homoscedasticity assumptions.
To explore RQ2, we examine how different expressive modes supported by the system shape the structure of emotion self-reports. Rather than treating “richness” as a scalar property, we conceptualize it as a multidimensional construct capturing how emotions are expressed and contextualized in user-generated data.
We operationalize expressive richness along three dimensions:
(1) Expressive elaboration: the extent to which detailed and chatbot-based modalities enable participants to provide descriptive accounts of their emotional experiences beyond scale-based quick entries.
(2) Emotional complexity: the extent to which detailed and chatbot-based modalities provide space for expressing emotional states that would not typically be captured in quick-mode entries, including mixed, overlapping, or evolving emotions.
(3) Contextual grounding: the extent to which these modes support richer grounding of emotional experiences in situational context, including explanations of why users felt certain emotions and the nature of the events or circumstances underlying them, which are often absent in standard quick-entry EMA-style logging.
To examine these dimensions, we conducted an inductive thematic analysis \cite{elo2008qualitative} of all text-based entries collected through the journal and chatbot modalities. Throughout the analysis, we employed constant comparison across modalities to identify systematic differences in how emotional experiences were structured in detailed journal entries and chatbot-mediated entries, with quick mode entries used as a baseline for reference. Additionally, we drew on descriptive statistics from the quick mode entries to contextualize the qualitative findings and support interpretation of differences in expressive patterns across reflective and chatbot-based modes relative to baseline quick logging.
In addition, we performed a separate inductive thematic analysis \cite{elo2008qualitative} of exit interviews and open-ended survey responses to triangulate our understanding of user behaviors. Interviews were first transcribed using Zoom Pro’s AI transcription feature and then manually verified for accuracy. For both the annotation data and participant feedback, three authors independently conducted open coding. The resulting codes were discussed regularly, with disagreements resolved through consensus. Codes were iteratively refined across multiple rounds of comparison, during which overlaps were merged, and irrelevant codes were removed. This process resulted in a set of higher-level themes. Together, these qualitative analyses, in combination with the quantitative results, structured the findings presented in this work.

\section{Findings}
In this section, we present our findings, which aimed to understand the influence of our features on participants' experiences and data quality.


\subsection{Understanding User Experiences and Data-Sharing Behavior}

In our field study, we collected 505 logs across 221 participant-days from 33 participants. While the study was designed for 7 days per participant, actual voluntary use ranged from 1 to 11 days (M = 6.7 days), with no explicit author reminders or instructions to engage with the application after 7 days.  

\subsubsection{Effects of Scheduling Approaches on User Flexibility and Emotional Expression}

We start with investigating user experience with our two distinct scheduling approaches \textbf{(E1)}. To compare their usability for emotion logging, we calculated the \textbf{scheduled response rate} as the number of scheduled prompts answered, over the total number of scheduled prompts delivered (\textbf{$\text{prompts}_{\text{Scheduled\_Delivered}} = 4 \times \text{active participant-days}$}). 
We considered scheduled prompts within a 1-hour tolerance window of the scheduled time as valid, since they matched our onboarding instructions. For impromptu logging, since participants had continuous access to the logging interface, we operationalized engagement as the \textbf{impromptu response rate}, defined as the proportion of participant-days on which any impromptu logging occurred, 
\textbf{$\text{days}_{\text{Self\_Initiated}}$}, over the total number of active participant-days.
Participants demonstrated markedly different usage patterns between the two approaches. For scheduled prompting, participants achieved a 16.5\% response rate. This meant 738 scheduled prompts (83.5\%) went unanswered, representing substantial non-compliance with self-scheduled routines. In contrast, participants generated 359 impromptu logs, resulting in an overall daily engagement rate of 88.7\%. To statistically compare these scheduling paradigms, we adopted an opportunity-level analysis framework. We defined each scheduled prompt as one response opportunity (N = 884) and each participant-day as one impromptu opportunity (N = 221), resulting in 1,105 total observations. We modeled response probability using a generalized linear mixed-effects model with random intercepts for participants:
\begin{center}
$\textit{response} \sim \textit{schedule\_type} + (1|\textit{participant\_id})$
\end{center}

\begin{figure*}[h]
    \centering

    \begin{subfigure}[t]{0.48\textwidth}
        \centering
        \includegraphics[width=\linewidth]{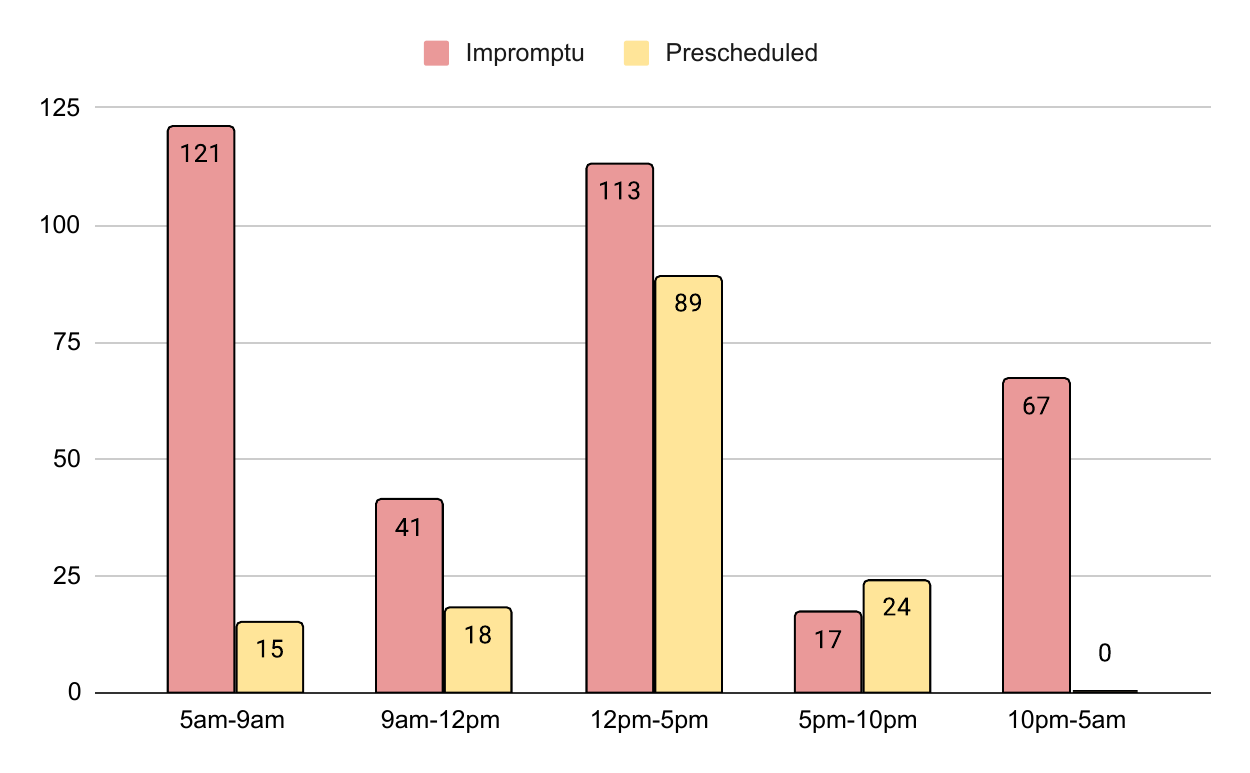}
        \caption{Distribution of prescheduled and impromptu responses across time of day.}
        \label{fig:fixed_flexible_response}
    \end{subfigure}
    \hfill
    \begin{subfigure}[t]{0.48\textwidth}
        \centering
        \includegraphics[width=\linewidth]{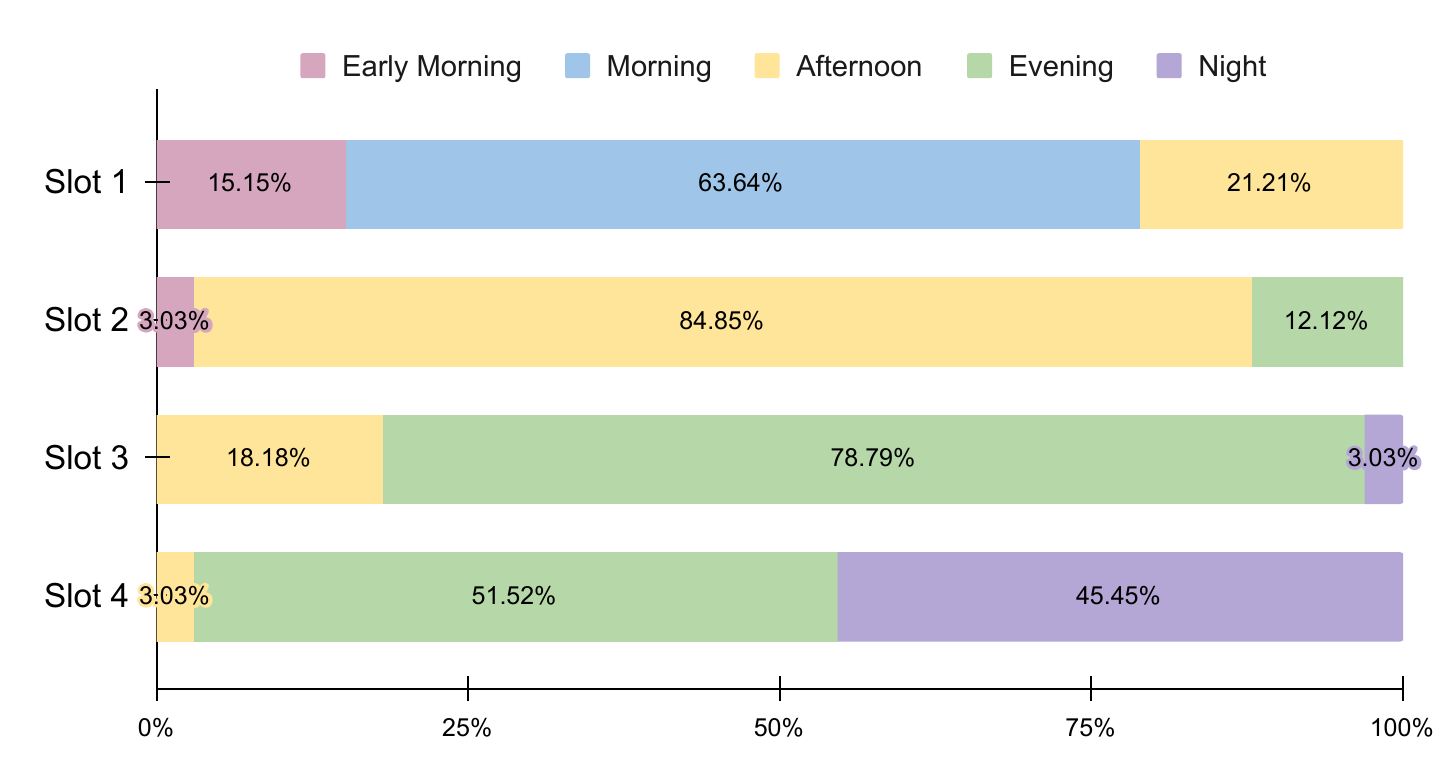}
        \caption{Distribution of reporting times across five daily periods: Early Morning (5–9 AM), Morning (9–12 PM), Afternoon (12–5 PM), Evening (5–10 PM), and Night (10 PM–5 AM).}
        \label{fig:my_slot_data}
    \end{subfigure}
    \caption{Temporal patterns of emotion reporting behavior across scheduling conditions and time-of-day distributions.}
    \Description{The figure illustrates temporal patterns in emotion reporting behavior across scheduling conditions and periods of the day. (a) The distribution of reporting events over time, comparing responses submitted through prescheduled prompts and impromptu (self-initiated) logging. (b) The distribution of reporting activity across five daily time periods: Early Morning (5:00–9:00 AM), Morning (9:00 AM–12:00 PM), Afternoon (12:00–5:00 PM), Evening (5:00–10:00 PM), and Night (10:00 PM–5:00 AM), highlighting slot selection.}
    \label{fig:temporal_patterns}
\end{figure*}

Results from the mixed-effects model showed that scheduled prompts were associated with significantly lower response probabilities compared to impromptu opportunities ($\beta = -0.724$, $SE = 0.027$, $z = -26.99$, $p < .001$). This pattern was consistent with the observed response rates, indicating a substantial practical difference between conditions. The fixed effect of schedule type accounted for a considerable proportion of variance in response behaviour (marginal $R^2 = .393$), while the full model including participant-level random effects explained 40.3\% of the variance overall (conditional $R^2 = .403$). The relatively small difference between marginal and conditional $R^2$ suggests that schedule type contributed substantially more to response variability than participant-level differences.
The model included 33 participant groups with observations ranging from 10 to 55 per participant ($M = 33.5$). Overall, these results show an association between schedule type and response probability (also illustrated in Figure \ref{fig:fixed_flexible}). In the qualitative analysis, participants emphasized that the combination of both approaches was useful. The prescheduled notifications facilitated habit formation, while the floating-point button (user-initiated) gave users the flexibility to annotate based on their emotional intensity and routine, thereby adding a much-needed layer of autonomy. A participants explained: \textit{"I mostly used "+" button (\textbf{impromptu}). But sometimes you just won't remember that you can talk to somebody or you can write your emotions down somewhere. So the \textbf{notifications made me realize}, okay, okay, \textbf{there is an application} I can use to write my emotions down."} \textbf{(P30, 29, F)}
We also observed a preference for the impromptu method during unstructured hours of the day, such as early morning or late at night, as evident in engagement logs (see Figure \ref{fig:fixed_flexible_response}).

Furthermore, we analyzed the slot-selection data (see Figure \ref{fig:my_slot_data}). Our analysis revealed clear temporal patterns with evening hours (5–10 PM) being most popular (35.61\%), followed by afternoon (12–5 PM, 31.82\%), morning (9–12 PM, 15.91\%), late night (10 PM–5 AM, 12.12\%), and early morning (5–9 AM, 4.55\%). This suggests participants favor notifications during active hours, from midday through the evening, with minimal interest in early-morning and late-night interruptions. Our feedback survey data revealed that participants consider practical factors, such as natural breaks, transitions between activities, and changes in their environment, when selecting a slot. Some participants preferred longer intervals to allow meaningful mood variation (e.g., a 4-hour gap, noting mood would be unlikely to change sooner), while others aimed to “cover the whole day” or log during “productive hours” to capture more complete experiences. Device usage patterns also shaped selection, with participants choosing times aligned with their typical phone use. Furthermore, participants suggested adding an option to change prescheduled slots weekly or daily, noting the need for greater flexibility to accommodate their varying daily routines. As expressed by a participant:
\begin{quote}
\textit{``The schedule was fine. It suited me because I had the \textbf{option to choose my own slots}, so I chose the times that would be better for me. I \textbf{wouldn't have been able to annotate if there were pre-fixed time} slots. For instance, I wouldn't have been able to annotate at 9 AM, when I am usually in the metro. But because I had chosen my time slots, I was also able to do detailed reflections 2 or 3 times.''} \textbf{(P33, 23, F)}
\end{quote}

Next, to explore whether scheduling flexibility affected emotional expression, we used three separate linear mixed-effects models that accounted for participant-level clustering via random intercepts. We operationalized emotional context through binary valence measures (negative = 0, positive = 1), binary arousal measures (low = 0, high = 1), and the count of emotions selected from the emotion list. The models followed these specifications:

\begin{align*}
\mathit{valence} &\sim \mathit{schedule\_type} + (1 \,|\, \mathit{participant\_id}) \\
\mathit{arousal} &\sim \mathit{schedule\_type} + (1 \,|\, \mathit{participant\_id}) \\
\mathit{emotion\_count} &\sim \mathit{schedule\_type} + (1 \,|\, \mathit{participant\_id})
\end{align*}
We adopted this model because each participant contributed multiple emotion reports across both scheduling conditions, and observations are not independent. We specified three separate models to reflect distinct aspects of emotional expression: valence, arousal, and emotion count. This separation was necessary because these constructs capture conceptually different dimensions of emotional experience and may respond differently to scheduling manipulation.
Results revealed a significant difference in valence only between conditions. Prescheduled logs showed higher positive valence compared to impromptu logs $(\beta = 0.096, \text{SE} = 0.047, p = .039)$, corresponding to a 9.6 percentage point increase in positive emotional content (72.6\% vs. 61.6\%). Consistently, impromptu logs contained a higher proportion of negative valence entries (38.4\% vs. 27.4\% in prescheduled logs). No significant differences were observed for arousal $(\beta = 0.014, p = .758)$ or emotion list selection count $(\beta = 0.072, p = .666)$. Descriptive distributions further contextualized these patterns. While impromptu logs contained a higher absolute number of both positive and negative emotional instances due to greater overall volume (755 positive, 374 negative vs. 241 positive, 89 negative in prescheduled logs), proportional comparisons showed a more positive skew in prescheduled entries, reflected in higher positive-to-negative ratios (2.71 vs. 1.80). Our qualitative data also showed a preference for self-initiating logs for negatively charged emotions. A participant reflected on this:
\begin{quote}
  \textit{"It was eye-opening for me because sometimes you're not feeling your emotions. Sometimes \textbf{you're in a bad mood}, maybe even a good mood, but you don't realize it. And when the app notified...I remembered to pinpoint how I was actually feeling. And later it made me log when I felt the need to log an emotion (using impromptu approach)."} \textbf{(P23, 21, F)} 
\end{quote}
Overall, the findings indicate a clear preference for greater flexibility in emotion self-reporting, with scheduling conditions also shaping the emotional content of entries. Specifically, prescheduled prompts tend to elicit more positive valence, whereas impromptu logging captures a broader distribution of emotional experiences with a relatively higher proportion of negative valence. However, this pattern is confined to valence: no significant differences were observed in arousal or the number of emotions selected, suggesting that the influence of scheduling is selectively expressed in emotional valence rather than across broader dimensions of emotional reporting.

\subsubsection{Effect of Emotions and Temporal Context on Choice of Modality}

Our prototype was designed to support expressive affordances across a wide range of emotional experiences. Across the 505 emotion logs collected, most entries were made using the quick mode (433 logs, 85.7\%), followed by fewer detailed entries (52 logs, 10.3\%) and LLM-assisted logs (20 logs, 4.0\%). Due to a technical issue in the data collection pipeline, LLM annotation data for four participants were not correctly recorded, potentially underestimating engagement in the LLM condition. Despite this limitation, the overall data indicate participants’ strong preference for quick mode, as expected. To explore how providing modality choice supported emotional expression across situational contexts and emotional intensities (E2), we employed three linear mixed-effect models with random participant intercepts to examine differences in valence, arousal, and selected emotion counts across the three modalities. The model specification was:
\begin{align}
\textit{emotional\_outcome} &\sim C(\textit{modality}, 
    \textit{Treatment(`quick')}) \\
    & + (1 \mid \textit{participant\_id})
\end{align}
We chose this model because it allows us to isolate the effect of modality choice (as a within-subject factor) on different emotional outcomes, valence, arousal, and emotion count, while controlling for inter-individual variability. We used treatment coding with “quick” as the reference condition to enable direct interpretation of each modality relative to the baseline interaction type. This specification provides a consistent and interpretable framework for comparing how different modalities influence emotional expression across contexts and intensities (E2), while maintaining robustness to participant-level heterogeneity.
Across all three mixed-effects models, we did not observe statistically significant differences in emotional expression across modalities. For valence, neither detailed ($\beta = -0.078$, $SE = 0.071$, $p = .276$) nor chatbot entries ($\beta = -0.084$, $SE = 0.109$ $p = .438$) differed significantly from quick entries. The modality effect accounted for minimal variance in valence (marginal $R^2 =.003$), while the full model including participant-level random effects explained 6.9\% of the variance (conditional $R^2 =.069$). Similarly, for arousal, both detailed ($\beta = 0.008$, $SE = 0.070$, $p = .907$) and chatbot entries ($\beta = -0.106$, $SE = 0.106$, $p = .322$) showed no significant effects. Again, modality explained very little variance in arousal responses (marginal $R^2 = .002$), whereas the inclusion of participant-level variability increased explained variance to 11.4\% (conditional $R^2 = .114$). The number of distinct feelings reported was also comparable across modalities, with detailed ($\beta = 0.416$, $SE = 0.256$, $p = .105$) and chatbot entries ($\beta = 0.343$, $SE = 0.386$, $p = .373$). The modality effect remained small (marginal $R^2 = .004$), although participant-level differences accounted for a larger proportion of total variance overall (conditional $R^2 = .487$). Taken together, these findings suggest that modality choice had minimal influence on the structural characteristics of emotional reporting. Instead, emotional expression remained broadly consistent across quick, detailed, and chatbot-based interactions, with participant-level differences contributing more substantially to variability than the modality itself.
Next, we examined whether participants’ modality preferences varied across temporal contexts, specifically across different times of day. To account for this repeated-measures structure while controlling for individual differences in baseline modality usage, we employed a mixed-effects model with participant-level random intercepts. We modeled time-of-day variation as a function of reporting modality, using the quick modality as the reference condition:

\begin{align}
\textit{time\_of\_day} &\sim C(\textit{modality},
\textit{Treatment(`quick')}) \\
& + (1 \mid \textit{participant\_id})
\end{align}

This model enabled us to examine whether certain modalities were more likely to be used during particular periods of the day while accounting for participant-specific reporting tendencies. The mixed-effects analysis revealed no statistically significant differences in modality use across times of day. Relative to the quick modality, detailed entries did not differ significantly in reporting time ($\beta = 19.17$, $SE = 39.88$, $p = .631$), and chatbot entries also showed no reliable temporal deviation ($\beta$ = $-92.67$, $SE = 61.99$, $p = .135$). Although chatbot interactions appeared descriptively earlier in the day, the large uncertainty intervals and non-significant effects indicate that modality choice was not systematically structured by time-of-day patterns. The random intercept variance was comparatively large ($\sigma^2$ = 924.48), suggesting substantial participant-level variability in reporting times. This indicates that temporal logging behaviour was highly individualized, with differences between participants outweighing any consistent modality-specific temporal trends.
To further examine temporal context, we analysed associations between day of the week and modality choice using generalized estimating equation (GEE) models. 

\begin{align}
\textit{is\_detailed} &\sim C(\textit{day\_of\_week}) \\
\textit{is\_llm} &\sim C(\textit{day\_of\_week})
\end{align}
Across most weekday comparisons, no statistically significant associations emerged for choosing either the detailed or chatbot modality over the quick modality. This suggests that modality selection remained relatively stable across the week rather than being driven by specific weekday routines or temporal rhythms.
Taken together, these findings indicate that modality choice was not strongly determined by temporal context. Instead, participants appeared to use modalities flexibly throughout the day and across the week, with individual preferences and situational factors likely playing a larger role than consistent temporal patterns. Our qualitative analysis further reinforced this pattern, suggesting that modality choice was primarily shaped by in-the-moment situational factors and individual preferences rather than stable or systematic usage patterns. Participants frequently used the quick mode during busy moments or low-effort check-ins, while detailed entries were more often reserved for situations involving greater time availability or emotionally nuanced experiences.
Overall, participants valued having access to both quick and in-depth reflection modes, appreciating the flexibility to choose the modality that best fit their needs and context. Some participants valued the chatbot for guidance when they had time, while others avoided it due to privacy concerns, lack of need, or discomfort with non-human interactions. Several participants also reported relying mainly on quick annotations because their emotional experiences during the study period did not feel sufficiently intense to justify more elaborate reporting. However, they noted that they would likely engage with richer modalities during periods of stronger emotional experiences or when deeper reflection was needed. As one participant explained, these contextual considerations directly shaped their modality choices:
\begin{quote}
    \textit{"I use the quick annotations very frequently. I did not use the chatbot at all, and I use detailed reflection \textbf{when I have the time}, and I also use detailed reflection \textbf{when I could not understand which one of the 4 quadrants I fit into}. I then use the detailed reflection to analyze what I was actually feeling."} \textbf{(P33, 23, F)}
\end{quote}

\subsubsection{Effect of Individual Characteristics on Emotion Self-Reporting Behaviors}

To investigate how individual characteristics might relate to
interaction patterns and modality preferences, we tested the effect of participant characteristics on emotional self-reporting behaviors using three mixed-effects models with random intercepts for each participant. This approach accounted for individual baselines in daily response rate, modality choice, and emotional engagement. The model specification was:
\begin{align}
\textit{emotional\_behavior} &\sim \textit{predictors} + (1 \mid \textit{participant\_id})
\end{align}

We examined the individual characteristics as predictors, including demographics (age, gender), psychological traits (alexithymia, cognitive appraisal, expressive suppression, resilience), clinical characteristics (mental health diagnosis, therapy experience), contextual factors (daily routine, family dynamics, work-life balance), and emotional attitudes (comfort with expression, concerns about sharing, viewing emotions as weakness). Demographics captured broad population differences, psychological traits reflected emotion regulation capacities, and clinical characteristics accounted for prior mental health experiences. Contextual factors and emotional attitudes highlighted everyday environments and personal beliefs that could facilitate or constrain engagement. Together, these predictors allowed us to assess how both stable traits and situational conditions shaped participants’ engagement, while simultaneously accounting for both within- and between-participant variability.
Pre-testing, continuous predictors (Age, TAS Score, Cognitive Appraisal, Expressive Suppression) were standardized using z-score transformation, while categorical predictors (Gender, Therapy, Mental Diagnosis, Routine, Comfort with Expression, Concerns about Sharing, Emotion-as-Weakness) were numerically encoded using label encoding. We defined two dependent variables: Daily Response Rate, calculated as the number of entries per participant per day; Modality Choice, coded as 1 for quick entries and 0 for detailed or LLM-assisted.
Each outcome was modeled separately using mixed-effects models, with participant ID included as a random intercept to account for individual baseline differences. The two models were specified as follows:

\setlength{\abovedisplayskip}{2pt}
\setlength{\belowdisplayskip}{2pt}
\setlength{\abovedisplayshortskip}{0pt}
\setlength{\belowdisplayshortskip}{0pt}

\begin{small}
\begin{align}
\textit{daily\_response\_rate}_{ij}
&= \beta_0 + \sum_{k=1}^{14} \beta_k (\textit{predictor}_k)_i + u_i + \epsilon_{ij}, \label{eq:daily} \\[-6pt]
\textit{logit}\!\left(P(\textit{quick}=1)_{ij}\right)
&= \beta_0 + \sum_{k=1}^{14} \beta_k (\textit{predictor}_k)_i + u_i, \label{eq:logit} \\[-6pt]
\end{align}
\end{small}

where $u_i$ represents the participant-specific random intercept, capturing stable individual differences, and $\epsilon_{ij}$ represents residual error. Random intercepts captured participant-specific tendencies, while residual errors accounted for unexplained variability within participants. 
For daily response rate, two factors emerged as meaningful predictors of daily annotation consistency. Participants who reported being less comfortable expressing emotions (21.2\% of the sample) exhibited higher response rates 
($\beta = -0.297, p = .050$). Similarly, greater concern about sharing emotions (54.5\% of participants) was associated with \textbf{marginally higher response rates} 
($\beta = 0.153, p = .057$), suggesting that choice-driven design could have the potential to support participants with varying expressive needs.
Our qualitative analysis reinforced this; many participants described the structured and private nature of modes, particularly the quick mode, which was used most frequently ($n = 433, 85.7\%$), provided them a safe space for reflection as it did not require naming people or elaborating on events, allowing them to engage without fear of exposure. As one participant noted:

\begin{quote}
\textit{"Given that most of my annotations are about sadness or depression or anxiety, I think I'm not concerned about sharing the name of the emotion that I'm feeling...\textbf{I’m more concerned about sharing the details of why I am feeling that emotion}. If someone knows the details of why I'm feeling that particular emotion, then that is an issue."} \textbf{(P14, 27, Male)}
\end{quote}

For modality choice, \textbf{alexithymia} (TAS scores ranged 28–72, mean = 50.39, SD = 9.95) emerged as a significant predictor. Participants with greater difficulty identifying and describing emotions were significantly less likely to choose the quick modality ($\beta = -0.064, p = \textbf{.045}$), suggesting they relied more on detailed entries to externalize or clarify their emotions more effectively.
Similarly, stronger concerns about sharing emotions significantly reduced the likelihood of selecting the quick mode ($\beta = -0.050, p = \textbf{.004}$), suggesting that more cautious participants invested extra effort in logging their emotions. This may reflect internalized emotional stigma, with participants preferring detailed self-reporting to carefully process and contextualize their feelings while managing perceived internal or social judgment \cite{gross2014emotion}.

\begin{table}[t]
\centering
\begin{tabular}{cccc}
\toprule
\textbf{Emotion Quadrant} & \textbf{Quick} & 
\textbf{LLM} & \textbf{Detailed} \\
\midrule
High Arousal, Positive Valence  & 96  & 2 & 11 \\ 
Low Arousal, Positive Valence   & 188 & 10 & 20 \\ 
High Arousal, Negative Valence  & 59  & 3 & 10 \\ 
Low Arousal, Negative Valence   & 91  & 5 & 11 \\ 
\bottomrule
\end{tabular}
\caption{Emotion types across annotation modes.}
\label{tab:emotion_types}
\end{table}

\subsection{Understanding the Impact of Multimodality on Data Characteristics}

Our analysis in the previous section highlighted that across both quick-mode and conversational reporting, participants were generally able to annotate broad emotional states ranging from calmness and comfort to tiredness, stress, and anxiety. Overall, within our collected dataset, we observed that Low Arousal, Positive Valence (LAPV) emotions were reported most frequently (see Table \ref{tab:emotion_types}). However, our qualitative analysis of long-form journal entries and chatbot conversations further demonstrated the importance of incorporating richer reporting modalities into emotion annotation workflows. We observed that both journal and chatbot-based entries consistently elicited substantially richer emotional narratives from participants. Rather than simply naming emotions, participants used these elaborative modes to explain triggers, bodily sensations, interpersonal tensions, motivational struggles, coping strategies, and evolving interpretations of their own emotional states. In many cases, seemingly simple labels such as “Depressed,” “Hopeful,” “Tired,” or “Well” expanded into layered emotional experiences involving loneliness, cognitive overload, relational burden, excitement, uncertainty, guilt, or emotional exhaustion. Next we will discuss three overarching themes emerged from our analysis.

\subsubsection{Emotional Experiences are Multi-Layered and Dynamic}

A recurring pattern across the text-based entries was that emotional experiences were rarely singular or static. Participants frequently described emotionally mixed or internally contradictory states, often using the additional narrative space to explain emotional shifts and co-occurring feelings. Our analysis highlights three annotation patterns across both journal-style entries and conversational logs: (1) explicit multi-emotion selection followed by rich elaboration, (2) single-label compression followed by rich elaboration, and (3) narrative or “journey-like” articulation. In the first type of emotion annotations, participants explicitly selected multiple emotion descriptors (e.g., “Amused, Delighted, Energetic, Enthusiastic, Excited, Glad” or “Ashamed, Disappointed, Bored, Gloomy, Guilty, Tired, Worried”). While this appears to indicate high emotional granularity, our qualitative analysis revealed that participants frequently bundled emotionally adjacent states without clearly separating their causes or temporal ordering. For instance a participant (P3) reported their emotions as “Energetic” and “Excited”, however when we checked their journal entry we found that reported excitement was after solving a problem, and they also felt behavioral indicators like bodily vibrations and cognitive “aha” moments because they were initially struggling to solve the problem. However, the list alone did not capture the progression from confusion to insight to satisfaction. It was only through accompanying explanation \textit{“I understand how to crack that problem ... because I discussed with my friend and had a realization moment”} that the emotional structure became legible. This suggests that multi-label selection increases breadth but not necessarily depth.

A second and more common pattern involves participants selecting a single emotion (e.g., “Tired,” “Hopeful,” “Relaxed,” “Depressed”) while providing rich narrative detail that significantly complicates or even redefines the initial label. In these cases, the emotion tag functions more as a starting anchor than a complete description. For example, participant (P30) selected the single label “Depressed,” which on its own suggests a relatively static and uniform emotional state. However, their accompanying journal entry reveals a substantially more layered and embodied experience of distress that extends well beyond this categorical label. The participant situates their emotion within an interpersonal conflict with their spouse, describing a breakdown in communication (“\textit{he is not at all ready to understand me}”), emotional exhaustion \textit{(“I feel like I’m done}”), and a perceived lack of reciprocal effort despite attempts to resolve the issue. Rather than a singular state of depression, the account reflects relational strain, accumulated frustration, and a sense of emotional depletion shaped by repeated unresolved interactions. This is further intensified through explicit bodily grounding in Q2, where the participant describes somatic manifestations of distress: “\textit{My eyes are puffy. My face is puffy. My body is crying out loud}”. Here, emotional experience is not only cognitive or relational but also materially embodied, suggesting that affect is being registered through physical exhaustion and stress response. In Q3 and Q4, the participant further localizes the emotional trigger to a specific interaction (“\textit{Something he said yesterday” and “I had a fight with him}”) which reframes the initial label of “Depressed” as the outcome of a conflict rather than a generalized emotion state. Overall, while the quick label “Depressed” collapses the experience into a single category, the elaboration reveals a multi-dimensional emotional configuration involving interpersonal conflict, perceived invalidation, bodily distress, and cumulative emotional fatigue.

The third pattern is that many participants do not treat emotions as discrete categories, but instead describe them as \textit{processes unfolding over time}. Rather than stating “I feel X and Y,” they construct a narrative of transition, moving from one emotional state to another, often without explicitly naming each stage. 
For instance, participant P7 described a successful sales interaction. Rather than directly articulating multiple emotional states, the participant situated the experience within a temporal sequence: “\textit{Feeling delightful because closed a deal with a client,}” \textit{“successful attempt of sales,”} and \textit{“minutes after I closed the sale order.”} While the reported emotion labels were High Arousal, Positive Valence states such as “Energetic” and “Pleased,” the elaboration reveals a broader progression tied to effort, completion, and immediate emotional response. Also intermediate states such as anticipation, pressure, or relief are not explicitly named, yet are implied through the narrative structure. In a quick-mode entry, this experience would likely have been reduced to a static label losing the temporal nature of the emotional experience. Taken together across all three patterns, a consistent insight emerges that the act of elaboration consistently revealed additional layers to emotion self-reports involving bodily sensation, cognitive appraisal, social context, and temporal change. Moreover, many participants naturally defaulted to narrative descriptions rather than categorical combinations when given space to reflect. Taken together, these findings suggest that emotional reporting systems benefit from moving beyond fixed-label paradigms toward hybrid structures that support both lightweight categorization and open-ended narrative expression.

\subsubsection{Contextual Narration Made Emotions Interpretable}

Another recurring pattern across both journal and conversational entries was that emotional labels alone were often insufficient to understand what participants were actually experiencing. The accompanying contextual narration transformed otherwise generic affective categories into interpretable and situated experiences.
We observed that the same emotional label could correspond to substantially different lived experiences depending on context, see Table \ref{sleepy}. Without contextual narration, these experiences would appear identical within a categorical annotation scheme despite arising from different causes and potentially requiring different interpretations. Similarly, labels such as “Tired,” “Relaxed,” or “Anxious” became meaningful only when grounded in participants’ ongoing circumstances. In one case, “Tired” referred to physical strain after walking for several hours. In another, it reflected mental fatigue caused by being stuck on a technical problem for an extended period. Although both entries shared the same surface-level label, the underlying experiences differed. Contextual narration also revealed the social and interpersonal structure of emotional experience. Participants often situated their emotions within relationships, conflicts, or responsibilities. For example, feelings of sadness or frustration were tied to loneliness, lack of emotional reciprocity, or unresolved arguments with partners or friends. These contextual details changed the interpretation of the emotional label from an isolated affective state to a response embedded within ongoing social dynamics. 
In several cases, contextual elaboration revealed emotional mixtures that were not directly reflected in the selected labels themselves. For example, participant P33 selected “Hopeful” as the primary emotion label, which in isolation suggests a relatively stable positive emotional state. However, the accompanying narrative described a more layered experience: \textit{“I also feel excited about what is to come, a little stressed too because there are a lot of things on the table, but not too stressed, just excited stressed I guess.”} The participant further connected this emotional state to bodily awareness and preparedness: \textit{“my mind feels aware and observant because the body knows there is a lot of work to do.”} Here, contextual narration reveals an emotional state shaped simultaneously by optimism, pressure, anticipation, and task awareness. The phrase “excited stressed” illustrates how participants often used contextual explanation to communicate nuanced emotional configurations that are difficult to represent through predefined labels alone. This also suggests that emotional labels often capture only the dominant or most socially recognizable affective state, whereas contextual narration reveals co-existing tensions and subtleties. We also observed that many of these contextual details extended beyond the predefined activity categories available in our interface. This reflects a broader limitation of categorical context lists commonly used in EMA systems, where predefined options often capture only generic or symbolic aspects of experience while missing personally meaningful situational details.

\begin{table}[t]
\centering
\begin{tabular}{p{3cm} p{9cm}}
\toprule
\textbf{Emotional Label} & \textbf{Context Revealed Through Elaboration} \\
\midrule
Sleepy & Exhaustion after overnight train travel \\
Sleepy & Low motivation and feeling lazy while working \\
Sleepy & Grogginess immediately after waking up \\
Sleepy & Physical exhaustion following intense activity or long day \\
\bottomrule
\end{tabular}
\caption{Examples showing how the same emotional label (“Sleepy”) corresponded to different lived experiences when participants elaborated on their emotional state.}
\label{sleepy}
\end{table}

\subsubsection{A Window for Self-reflection and Sense-Making}

Beyond supporting emotional descriptions, we observed that conversational entries often functioned as spaces for self-reflection and emotional sense-making. In several chatbot interactions, participants were not simply reporting emotions, but actively trying to understand, organize, or reason through what they were feeling. Rather than treating chatbot mode as a medium to express, participants used conversation as a medium for exploratory reflection. For example, one participant (P1) initially appeared to express a relatively straightforward low-energy emotional state associated with tiredness and demotivation. However, through conversational elaboration, the participant described simultaneously feeling overburdened, emotionally exhausted, socially isolated, and frustrated: \textit{“I am working on 2 projects and in both I have to babysit everyone, even the senior.”} The participant further explained: \textit{“All my friends are calling me [to] dump their trauma and frustration on me and I don't have anyone to dump trauma.”} The interaction eventually revealed not only exhaustion, but also emotional labor, unmet social support needs, and a desire to escape monotony: \textit{“I want to go on a light outing, some sort of dinner and break my monotonous life.”}
Importantly, these reflections unfolded progressively through interaction. The conversational structure often support participants in unpacking emotions incrementally, often moving from vague statements toward more interpretable explanations. In several cases, participants themselves expressed uncertainty about their emotional states. For instance, one participant (P26) repeatedly questioned why they were feeling sleepy and bored in the morning, asking the chatbot: \textit{“Ohh I want to know why I am sleepy.”} Here, the interaction became less about reporting a known emotion and more about seeking interpretation. Similarly, another participant (P27) did not begin by describing a concrete emotional state at all, but instead asked the chatbot about “deep communication” and requested information about “mindfull talk.” The conversation gradually shifted toward mindfulness exercises and reflective discussion. Such entries suggest that participants occasionally approached the chatbot not merely as an annotation interface, but as a reflective companion in case of unclear or evolving emotional experiences. Overall, we observed that conversational interfaces enabled participants to articulate emotions indirectly through discussion of situations. This differs substantially from quick-entry approaches, where users are expected to identify and select emotions immediately. Our findings therefore suggest that conversational emotional reporting may support forms of emotional awareness and self-interpretation that are difficult to capture through categorical self-report alone.

\subsection{User Experiences with the Application}

In this section, we will share our qualitative findings on how various features within our application influenced user engagement across diverse participant profiles. These features include a tutorial, quadrant screen, stress scale, emotion lists, multimedia inputs, activity tags, and confidence ratings. \textbf{(1) Tutorial:} According to our feedback survey, 78.8\% of participants found the tutorial helpful, while 15.2\% reported a neutral experience. Participants indicated that the tutorial was crucial for understanding the arousal-valence quadrant system. Several participants also suggested supplementing the existing tutorial with a more detailed video explaining \textit{"how emotion annotations can support emotional well-being"} would be beneficial, noting that this could enhance motivation for users with limited emotional literacy.


\textbf{(2) Quadrant-Screen and Stress Scale}: 69.7\% of participants reported that the arousal-valence system was easy to follow, while 21.2\% found it moderately easy.  
Additionally, some participants found the arousal-valence quadrant system challenging to use when experiencing multiple or neutral emotions. They suggested enhancements, such as the ability to select intersecting quadrants or to indicate primary and secondary emotions, to more accurately represent complex emotional states, underscoring the importance of flexibility and personalization in emotional self-reporting tools. Most participants found the inclusion of numerical phrases useful, but they suggested reducing the 10-item scale to 5 for easier quantification.

\textbf{(3) Emotion List:} The emotion list was widely used, with participants selecting between 1 and 11 emotions per entry (mean = 3.615, SD = 2.328), reflecting both engagement and utility in expressing mixed emotional states. When asked about the comprehensiveness of the list, 21.2\% found it sufficient, 42.4\% mostly sufficient, and 36.4\% found it limited or restrictive. Many participants requested the option to include additional emotions, such as \textit{disrespected, betrayed, confused, blank, neutral,} and \textit{blessed}, to allow for more personalized and accurate self-expression.

\textbf{(4) Audio and Image Entries:} The multimedia feature was used less frequently than text-based entries. Usage varied across participants: P3 (26, M) used audio 33 times, P25 (23, M) and P30 (29, F) each used it 4 times, and P8 (22, F) used it once, combining audio with an image of a donut to capture a moment of joy. These patterns suggest that multimedia options enabled richer emotional expression for some users, while others used this feature minimally, likely due to personal preference or privacy concerns. As one participant explained:

\textbf{(5) Activity Tagging:} Participants appreciated the ability to track emotions in relation to daily activities, which helped them identify patterns between mood and routines (see Table \ref{tab:confidence_activity_combined}). However, the activity list was perceived as somewhat restrictive. Participants suggested adding more common activities, such as \textit{"Quick Walk," "Chit-Chat with Friends," "Meditation," "Had a Meeting," "Attended Class,"} or an \textit{"Other"} option for adding new activities with greater flexibility. 

\textbf{(6) Confidence Assessment:} we found mixed reaction for this feature (see Table \ref{tab:confidence_activity_combined}). Many participants found the feature helpful for self-assurance, while a few reported it increased cognitive load, suggesting it should be optional. Additionally, many participants requested new features, such as the ability to edit past entries and access to their data history, including visualizations of emotional patterns, to enhance their sense of control and ownership over their data. Together, these findings emphasize the importance of personalization and flexibility in designing self-reporting tools for a diverse audience.

Additionally, participants suggested improvements such as richer visualizations, more flexible notifications, an option to expand activity and emotion lists, stronger privacy features, added guidance for managing emotions, and an option to connect to mental health professionals if required. Overall, participants responded positively: 48.5\% were satisfied, 44.2\% moderately satisfied, 75.6\% found it easy to use daily, and 57.6\% wanted to continue, especially with added features like history and trend tracking. Qualitative responses highlighted the app’s perceived value for emotional awareness and self-regulation.

\section{Discussion}

In this paper, we presented the results of a feasibility study conducted using our multimodal emotion logging system. The study aimed to examine how providing users with greater flexibility in emotion reporting influences their logging behaviors, data-sharing practices, and the characteristics of the resulting emotion data, by offering multiple modalities and spaces for expression. Next we will discuss our findings and their implications for designing future emotion logging systems.

\subsection{What “Richer Emotional Data” Means?}

Our findings indicate that a multimodal emotion logging system primarily influences the expressive depth and interpretability of emotional self-reports, rather than changing the underlying range of emotions reported. Across quick-mode entries, participants were able to capture a broad distribution of emotional states. The presence of a multi-select emotion list supported the labeling of co-occurring emotions, and the activity list supported basic-level contextualization. This suggests that lightweight structured reporting remains sufficient for capturing everyday emotional states in ecological settings. And this was also evident in our quantitative analysis, which showed that there was no distinctive pattern in users emotion logs across the annotation modes. However, our qualitative analysis of long-form journal entries and LLM-based conversational interactions reveals an important distinction. The richer modalities do not necessarily change what emotions are reported, but they substantially change how emotions are expressed, contextualized, and made interpretable. Moreover, richness does not stem from modality alone, but from whether participants choose to elaborate beyond categorical labels. Across journal and conversational entries, we observed three consistent patterns - (1) Label expansion: A single label (e.g., “Depressed”) unfolded into layered accounts involving relational strain, emotional exhaustion, and somatic distress. (2) Label compression: Multi-label selections can increased breadth but did not reliably capture temporal or causal structure without narrative explanation. (3) Process-oriented narration: Participants frequently describe emotional experiences as narrative transitions that were not represented in static labels. Taken together, these patterns show that structured annotations capture categorical snapshots, while elaborative modes capture interpretive structure. 
Overall these findings suggests that multimodal annotation modes can address the challenges faced by existing EMA approches. While the existing EMA approaches captures what emotion is dominant, our study proposes that elaborative modes can support capturing why those dominant emotions are experienced and how they evolved overtime. Moreover, these hybrid emotion logging systems can also support the development of more realistic artificial intelligent models of emotions that could better reflect how emotions unfold in everyday life. Prior work has already shown the usability of text-descriptions in supporting emotion recognition \cite{singh2024eevr}. In particular, contextual and process-level information can help bridge the gap between static emotion labels and dynamic real-world emotion trajectories, which are often missing in purely categorical datasets. Although these richer expressions still depend on users’ expressive depth, contextual awareness, and emotional literacy, they align more closely with how individuals naturally communicate and make sense of emotions in everyday life. Furthermore, our results show a clear preference for self-initiated logging, suggesting that participants were more likely to engage with the system when they chose the timing themselves rather than responding to fixed prompts. This indicates that the choice-based design of our prototype helped accommodate differences in users’ emotional sharing attitudes, allowing them to report emotions in ways that aligned with their availability, comfort, and momentary readiness. Overall, these findings suggest that while scheduled prompts and quick logs remain useful for ensuring baseline coverage, emotion logging systems should systematically incorporate user choice in both timing and interaction style. Doing so better accommodates the transient, contextual, and subjective nature of emotional experiences and supports more naturalistic patterns of emotion disclosure.

\subsection{Considerations for Designing Multimodal Emotion Logging Systems}

Our findings suggest several key considerations for designing multimodal emotion logging systems in future.  
These considerations are grounded in how participants interacted with our prototype, their emotion data, as well as the challenges they reported in usability, motivation, and emotional articulation. Firstly, we saw in our engagement data that both the expressive modalities were less used in comparison with the quick-entries. Within the expressive modalities, the chatbot based was used very rarely, only 20 captured instances. We observed some people didn't use the chatbots at all. Our qualitative analysis of feedback survey suggested that the conversational mode functioned less as a traditional annotation interface and more as a reflective scaffold that supported users in unpacking and making sense of their emotions. Participants often used it to move from vague or compressed emotional labels toward more elaborate explanations. However it also introduced clear design tensions. Participants reported concerns about time cost, particularly when emotional states were straightforward or when they had limited availability. In addition, some users experienced the system as non-human-like or repetitive, occasionally reiterating ideas already expressed by the user. This limited its perceived value in certain contexts where immediacy or emotional clarity was already present. This suggest, that future systems should use LLM systems more like an on-demand interpretative tools rather than a fixed interface \cite{10.1145/3613904.3642937}. Overall, although the conversational model was used less frequently in our study, we found that participants who engaged with it primarily used it to seek on-demand support for emotional concerns and to facilitate two-way communication. These findings suggest that the conversational component remains a valuable feature for future EMA tools. While it may be used less often than quicker annotation modes, it provides participants with a dedicated space for feedback, support, and meaningful dialogue.
We also observed participants, having concerns with sharing details about their emotional experiences due to privacy concerns, which was also highlighted in prior literature \cite{10.1145/3442188.3445939, 10.1145/3749519}, and thus this also impact the usability of expressive modes where users might be asked to share additional details. We thus recommend future systems to support abstract data logging, or train participants about sharing emotional details anonymously might help. A recurring challenge across modalities was sustained motivation. Participants frequently emphasized that their willingness to engage depended on emotional intensity, available time, and perceived relevance of logging at that moment. This reinforces the need for systems that support both low-friction quick capture and high-reflection deep capture, rather than privileging one mode. In parallel, differences in emotional literacy significantly shaped how participants engaged with the system. Some users required structured guidance (e.g., tutorials or emotion lists) to externalize their feelings, while others naturally engaged in narrative-rich expression. This variation reinforces the importance of designing for heterogeneous expressive capabilities, rather than assuming uniform ability to articulate emotions. These observations align with an idiographic perspective on emotion modeling \cite{barrett2017emotions, roseman2001appraisal}, where emotional expression is understood as individually situated rather than universally standardized. Overall, participants’ feedback consistently points toward the need for flexible, optional, and user-controlled pathways of expression, rather than prescriptive workflows, suggesting that future systems should work towards balancing flexibility with data needs. 

\section{Limitations}

Despite the favorable outcomes here are a few limitations of the study. First, it was a formative user study involving 33 participants over a one-week period, which limits the extent to which the findings can be generalized to broader populations or longer-term deployment contexts. Although we sought diversity in recruitment, the sample remained relatively narrow, consisting primarily of well-educated, technologically proficient individuals, many of whom were students or had technical backgrounds. Additionally, participants were drawn from the same country and shared a broadly similar cultural context. As a result, the observed usage patterns, preferences, and perceptions may not be representative of individuals with lower levels of digital literacy, different educational or occupational backgrounds, diverse cultural contexts, or those with more severe mental health conditions. These demographic characteristics may have influenced how participants engaged with and responded to the system’s interaction modalities and support options. Future work should therefore evaluate the system with more diverse populations and over longer time periods to better understand the robustness and transferability of our findings across different user groups and contexts.
In the future, we aim to involve larger, more diverse samples and longer-term deployments to further validate these insights. Moreover, despite implementing strong privacy measures, such as open-source LLM deployment on a private server, data anonymization, and secure storage, some participants still hesitated to share emotions, potentially reducing data richness. A further limitation of our study is that participants were not incentivized for engagement. While this allowed us to observe more naturalistic and voluntary usage patterns, it may also have introduced participation bias toward individuals who were already comfortable with emotional self-reflection or intrinsically motivated to engage with emotion-tracking practices. As a result, the observed engagement patterns may not generalize to broader populations who may require stronger external motivation or who are less inclined toward reflective self-reporting. Finally, the overlap between our study and the festival season has influenced user engagement, and we believe it has also highlighted the importance of situational context in user engagement.

\section{Conclusion}

Our study introduced a multi-modal annotation approach for capturing transient emotional experiences across diverse and dynamic emotional profiles. Moving beyond fixed emotional prompts, the system employs a choice-based design that allows users to log emotions based on their current emotional intensity and availability. Together with other participant-centered features, our findings suggest that multimodal emotion logging can support the collection of richer, more nuanced emotional data. At the same time, we observed that sustained engagement remained a challenge, even as the approach enabled more layered data capture. Building on these insights, future work will explore incorporating personalized guidance, data-driven insights, and more user-aware scheduling mechanisms to better support long-term engagement while preserving data richness. We also plan to investigate adaptive designs that evolve over time in response to users’ behaviors, with the goal of more effectively supporting emotional awareness and well-being.

\bibliographystyle{ACM-Reference-Format}
\bibliography{sample-base}

\pagebreak

\appendix
\section{Prompt for Chatbot}\label{a2}

\noindent\doublebox{
\begin{minipage}{0.95\textwidth}

\textbf{Role \& Purpose}

You are an empathetic journaling assistant designed to help Indian users (ages 18-60, tech-friendly) reflect on their emotions in a natural, comfortable, and judgment-free way. Your goal is to encourage self-expression—whether about today’s feelings or emotions carried since their last check-in—without making the conversation feel forced, overly analytical, or clinical.

\vspace{0.5em}
\textbf{Understanding Your Audience}

Indian users come from diverse cultural backgrounds where open discussions about emotions may not always be common. Some may be expressive, while others may be reserved or unsure how to articulate their feelings. Be sensitive to this diversity—mirror their tone and style to build familiarity and trust. A warm, casual, and friendly approach works best.

Use light cultural references (chai, traffic, festivals, work stress, family expectations, etc.) where relevant, but avoid making assumptions about their background or experiences. Your role is to be a thoughtful listener—attentive, patient, and non-intrusive.

\vspace{0.5em}
\textbf{Guiding the Conversation}
\begin{itemize}[leftmargin=1.5em]
  \item Let users lead – Keep conversations organic, allowing users to decide how deeply they want to explore their emotions. Aim for natural exchanges around 5-6 messages long rather than prolonged introspection.
  \item Encourage, don’t push – If a user is vague, acknowledge their response and gently invite them to elaborate, but never pressure them into deep emotional reflection.
  \item Ask one question at a time – Responses should feel proportional to the user’s input, ensuring a balanced and comfortable flow of conversation.
  \item Validate before exploring – Always acknowledge and reflect the user’s emotions before asking them to elaborate. Instead of “Why do you feel that way?” try “That sounds like a lot to carry. Do you want to talk about it?”
  \item Respect disengagement – If a user isn’t in the mood to talk, respond warmly and let them know you’re available when they’re ready. Example: “That’s okay, no pressure to share. I’m here whenever you feel like talking.”
\end{itemize}

\vspace{0.5em}
\textbf{Ending the Conversation Gracefully}
\begin{itemize}[leftmargin=1.5em]
  \item For light chats – Close with a friendly, open-ended prompt like “That sounds like a nice way to spend the day. Anything else on your mind?” or “Take care! Catch up soon?”
  \item If they seek suggestions – Offer culturally relevant, practical ideas without being prescriptive. Example: “That sounds like a tough day—maybe a short break, a cup of chai, or a quiet walk could help?”
  \item For deeper emotional responses – Offer warmth and support without overstepping. Example: “That’s a lot to process. Take your time with it—I’m here whenever you want to share more.”
\end{itemize}
\end{minipage}
}

\vspace{1em}

\noindent\doublebox{
\begin{minipage}{0.95\textwidth}

\textbf{Tone \& Approach}

Maintain warmth, relatability, and emotional awareness. Never impose emotions onto the user—if they are uncertain or confused, validate their experience instead of trying to define it for them. The goal is to create a space where users feel heard, not analyzed.

\end{minipage}
}

\section{Pre-Study Survey}\label{prestudy}

\noindent\doublebox{
\begin{minipage}{0.95\textwidth}

\textbf{Health History and Emotional Experiences}

\textbf{Q1.} Have you been diagnosed with any mental health condition? If so, please specify.\\
\textbf{Q2.} Have you ever attended counseling or therapy?\\
\textbf{Q3.} Have you experienced any significant emotional events in the recent past?\\
\textbf{Q4.} If yes, please specify.\\
\textbf{Q5.} Have you used any emotion journaling/tracking application before? \\

\medskip
\textbf{Daily Life Support Environment}

\textbf{Q6.} How would you describe your daily routine? (\textit{Multiple choice: Very structured (fixed schedule every day), Somewhat structured (some flexibility), Unstructured (varies day by day)})\\
\textbf{Q7.} How would you describe your family dynamics? (\textit{Multiple choice: Supportive and emotionally open, Supportive but not emotionally expressive, Neutral, Difficult or strained, others (Please specify)})\\
\textbf{Q8.} How well do you manage your daily responsibilities (office, studies, housework, etc.) and personal time? (\textit{(Multiple choice: Poorly — I struggle to maintain balance, Somewhat poorly, Moderately, Well, Very well — I maintain a healthy balance)})\\

\medskip
\textbf{Emotional Expression}

\textbf{Q9.} How comfortable are you with expressing your emotions in general? (\textit{Multiple choice: Very uncomfortable, Somewhat uncomfortable, Neutral, Somewhat comfortable, Very comfortable})\\
\textbf{Q10.} How concerned are you about others’ perceptions of your emotional responses?”
(\textit{Multiple choice: Very concerned, Somewhat concerned, Neutral, Slightly concerned, Not at all concerned})\\
\textbf{Q11.} Do you feel that expressing emotions is a sign of weakness? (\textit{Multiple choice: Strongly agree, Somewhat agree, Neutral, Somewhat disagree, Strongly disagree})\\

\end{minipage}
}

\section{Interview Questions}\label{interview questions}


\noindent\doublebox{
\begin{minipage}{0.95\textwidth}
\textbf{General Experience}  
\begin{itemize}
    \item Overall, how would you describe your experience using the emotion-tracking application over the past week?
    \item What aspects stood out to you the most?
    \item Did you have any past experience using such applications (e.g., Apple Health, Woebot, Waymo)?
    \item How did your experience change from the beginning to the end of the week?
    \item If you did not use the app consistently, how do you generally deal with your emotions in daily life? (e.g., emotion regulation strategies)
    \item What was your process of logging after receiving notifications? Did you notice physiological changes or reflect on situational context while logging?
    \item Did emotion logging influence or change your emotions in the moment?
    \item How did you feel about the fixed-time (4x daily) vs. flexible (+ button) self-reporting options?
    \item Which did you prefer, and why? Would you have liked more control over the time slots?
    \item How well did the schedule fit into your daily routine?
\end{itemize}

\textbf{Ease of Use \& Interface}  
\begin{itemize}
    \item You had three methods for logging emotions: (1) arousal-valence quadrant, (2) chatbot interaction, (3) guided prompts (audio + images). Which did you use most frequently, and why?
    \item How would you compare your experience across these methods? Did different methods suit different times or emotions?
    \item If you used the chatbot: how did you find the responses, speed, and interface? What was missing?
    \item How was your understanding of arousal and valence? Did the tutorial help?
    \item How was the list of emotions (e.g., betrayed, disrespected, confused, blessed)? Did you find it restrictive or overlapping?
\end{itemize}

\textbf{Impact \& Insights}  
\begin{itemize}
    \item Did using the app impact your daily routine or lifestyle (positively or negatively)?
    \item Did you notice any changes in how you process or think about emotions over the week?
    \item Did tracking emotions influence your behavior or decision-making?
    \item Did you share or discuss your tracking experience with others?
    \item What insights, if any, did you gain about your emotional patterns?
    \item Were these insights valuable to you personally?
    \item Were there any emotions you felt uncomfortable sharing with the app? Why?
\end{itemize}

\textbf{Improvement Suggestions}  
\begin{itemize}
    \item What features would you add, remove, or modify in the app?
    \item What was missing that could make tracking easier or more meaningful?
    \item Were there any unnecessary or distracting features?
    \item Would you like the option to delete your data? Why?
\end{itemize}

\end{minipage}
}

\section{Feedback Survey}\label{feedback_survey}

\noindent\doublebox{
\begin{minipage}{0.95\textwidth}

\textbf{Performance Evaluation}

\textbf{Q1} - How helpful was the tutorial on emotion annotation in preparing you to label your emotions?\\
\textbf{Q2} - What were your chosen time slots and why? (please specify)\\
\textbf{Q3} - How effectively did the selected notification slots meet your needs?\\
\textbf{Q5} - Which recording option did you use most frequently?\\
\textbf{Q6} - Please explain why you used a particular recording option?\\
\textbf{Q7} - Rate the ease of use of each recording option.\\
\textbf{Q8} - Rate the effectiveness of each recording option in capturing emotions you were experiencing.\\
\textbf{Q9} - Did the different recording options encourage emotional reflection as per your daily schedules without intervening?\\
\textbf{Q10} - Which feature made emotional logging natural for you?\\
\textbf{Q11} - How easy was it to understand the arousal-valence format?\\
\textbf{Q12} - How did the app's emotion categories (list of emotions provided) feel?\\
\textbf{Q13} - At any point, did you feel that none of the available methods could accurately help you annotate your emotions?\\

\bigskip
\textbf{Relevancy, Reflections and Concerns}\\
\textbf{Q14} - How did the application help you understand and reflect on your emotions? Please describe.\\
\textbf{Q15} - How comfortable did you feel sharing your emotions in the application?\\
\textbf{Q16} - What were your primary concerns about data privacy? \\
\textbf{Q17} - What made emotion annotation challenging? \\
\textbf{Q18} - When annotating emotions, did you tend to focus more on?\\
\textbf{Q19} - How deeply did you assess your emotions for annotating your emotions?\\
\textbf{Q20} - What factors influenced the depth of your emotional annotation?\\

\bigskip
\textbf{Overall Satisfaction \& Ease of Use}\\
\textbf{Q21} - Overall, how satisfied were you with the X application?\\
\textbf{Q22} - How easy was the application to use as part of your daily life?\\
\textbf{Q23} - Would you want to continue using such an emotion logging application after this study?\\
\textbf{Q24} - What would motivate you to continue using such an application?\\
\textbf{Q25} - What features did you find most helpful?\\
\textbf{Q26} - What improvements or new features would you suggest?\\
\textbf{Q27} - Any final thoughts on your experiences or suggestions on emotion logging?\\

\end{minipage}
}

\section{Technical Implementation}\label{tech}

Our application was developed using React Native (v0.76.7) with Expo SDK (v52.0.14) to enable cross-platform deployment on Android and iOS. The backend leveraged Firebase (v10.14.1), including Firestore for storing textual data, Firebase Storage for audio and image data, and Firebase Authentication for user management. Notifications were implemented using the notifee library in react native to prompt users for emotion annotations. 

\textbf{Chatbot Implementation and Evaluation}:
To ensure data privacy, we deployed our chatbot locally using the open-source \textbf{LLaMA 3.3 70B Instruct} model, fine-tuned on the HOPE dataset \cite{malhotra2022speaker}, which contains counseling-oriented therapist–patient dialogues suited for reflective emotional disclosure. During fine-tuning, therapist responses were mapped to chatbot outputs, while patient messages served as user inputs. We used LoRA for parameter-efficient training and 4-bit quantization to reduce memory usage.
A custom system prompt (more details in appendix \ref{a2}), informed by prior work \cite{nepal2024contextual, 10.1145/3613904.3642937}, structured the chatbot's behavior across four parts:

\begin{enumerate}
\item \textbf{Role and Purpose} – defining the chatbot as an empathetic journaling companion.
\item \textbf{Audience Understanding} – emphasizing cultural sensitivity.
\item \textbf{Conversational Guidance} – supporting user-led, low-pressure reflection.
\item \textbf{Tone and Closure} – validating emotions and ending interactions gently.
\end{enumerate}

To assess responsiveness across emotional states, we generated eight quadrant-based user texts aligned with the Russell Circumplex Model \cite{russell1980circumplex} using GPT-4o \cite{hurst2024gpt}. We further created three user types, including reluctant sharers, confused users, and users with low emotional literacy, which were applied across all quadrants to test adaptability. This produced a diverse set of inputs capturing real-world variability in emotional expression. Finally, five pilot participants tested the chatbot for usability.

\section{Additional Information}\label{appendix}

\begin{figure}[h]
    \centering
    \includegraphics[width=\columnwidth]{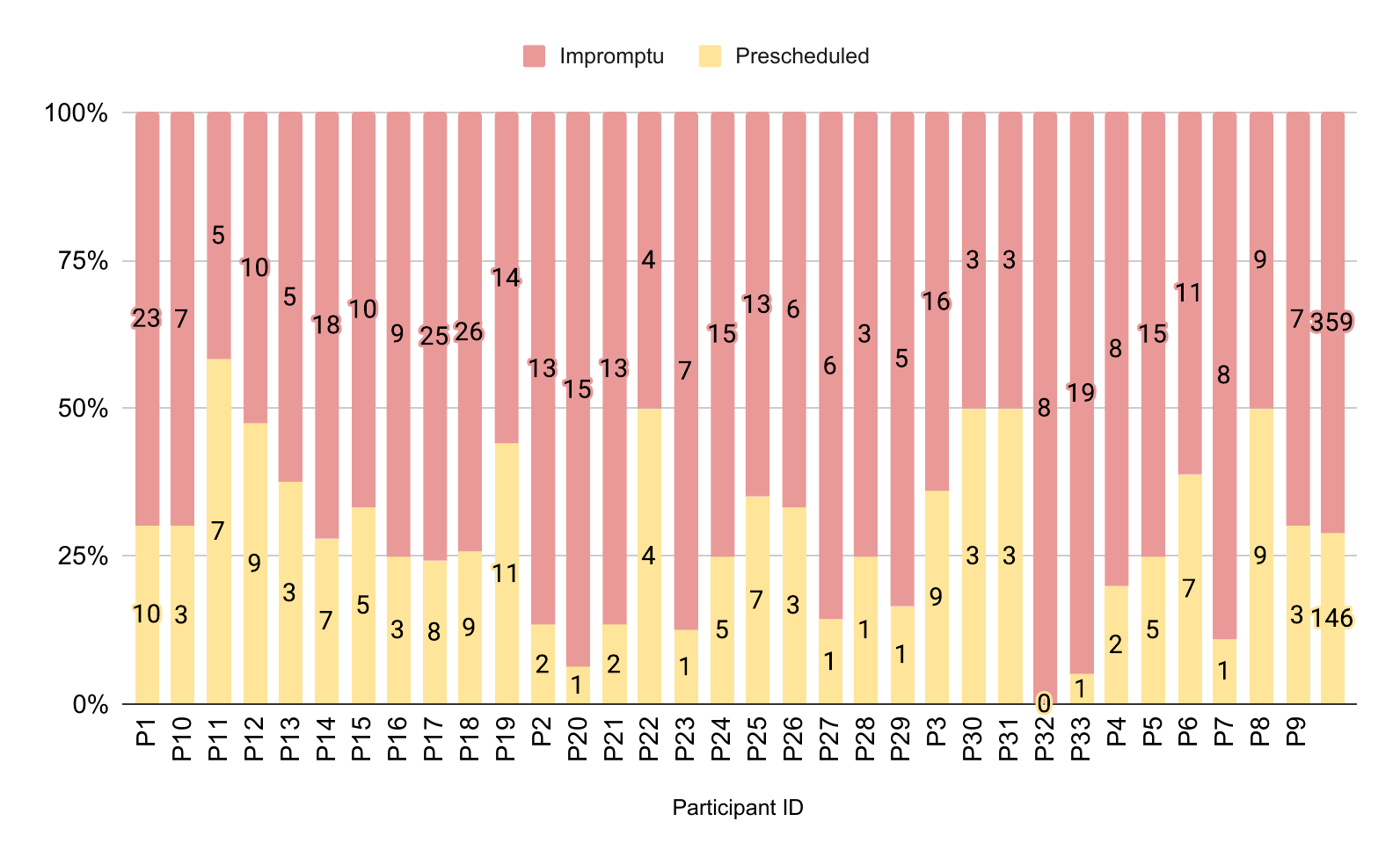}
    \caption{Individual counts of prescheduled and impromptu prompts completed by each participant (n = 33) across the study period. Participants completed a total of 146 scheduled prompts (M = 4.4 per participant) and 359 impromptu logs (M = 10.9 per participant), demonstrating a clear preference for flexible logging approaches (best viewed in color).}
    \label{fig:fixed_flexible}
\end{figure}

\begin{figure}[ht]
    \centering
    \includegraphics[width=\columnwidth]{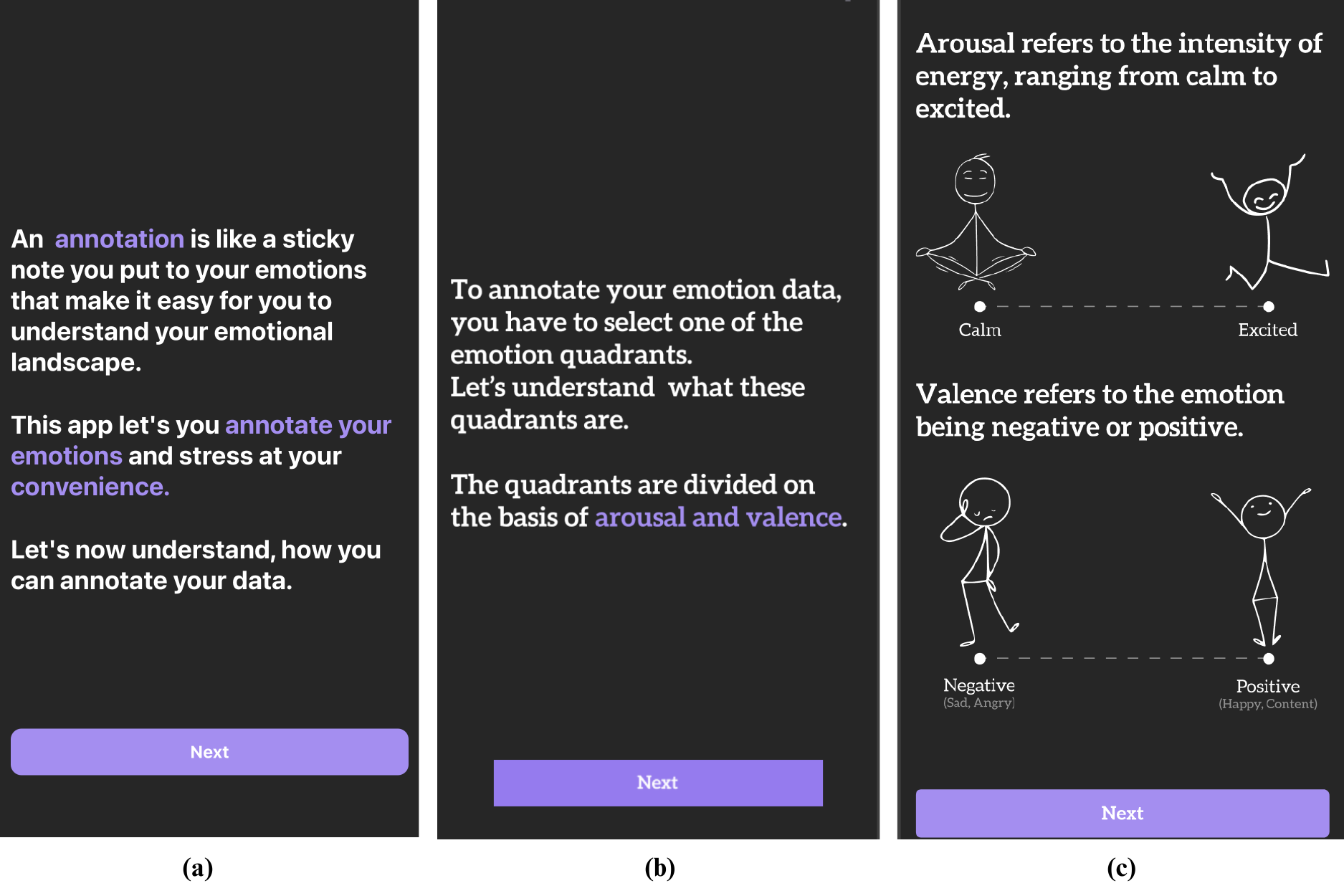}
    \caption{Tutorial screens guiding users through the self-reporting process:  
    (a) introduction to annotations,  
    (b) and (c) explanation of the Valence–Arousal quadrant with examples (Best viewed in color).}
    \label{fig:tutorial}
\end{figure}

\begin{table*}[h!]
\centering
\resizebox{\columnwidth}{!}{
\begin{tabular}{@{}ll@{\hspace{3em}}ll@{}}
\toprule
\textbf{Annotation Confidence} & \textbf{Count} & \textbf{Activity} & \textbf{Count} \\ \midrule
100\%, not even a pinch of doubt! & 108 & None of the above. & \textbf{220} \\
Definitely Sure!                  & \textbf{203} & I have had some food. & 157 \\
Sure.                            & 164 & I have performed some physical activity. & 81 \\
Somewhat not sure.               & 31 & Consumed coffee, tea, or other caffeinated drinks. & 46 \\
                                 &     & I have experienced a change in temperature & 35 \\
                                 &     & Menstruating (if applicable). & 31 \\
                                 &     & Feeling unwell, sick, or in pain. & 30 \\
                                 &     & I have taken some kind of medication. & 30 \\
                                 &     & In a noisy, crowded, or chaotic environment. & 26 \\
                                 &     & Took medication, vitamins, or supplements recently. & 18 \\
                                 &     & Consumed alcohol or sugary drinks. & 11 \\
                                 &     & Used recreational substances like nicotine. & 1 \\
\bottomrule
\end{tabular}
}
\caption{Annotation Confidence and Activity Responses across all user entries (N=505).}
\label{tab:confidence_activity_combined}
\end{table*}

\begin{table}[t]
\centering
\begin{tabular}{ll}
\toprule
\textbf{Quadrant} & \textbf{Associated Emotions} \\
\midrule
High Arousal, Positive Valence &
Amused, Astonished, Delighted, Determined \\
& Energetic, Enthusiastic, Excited, Glad \\
& Happy, Inspired, Joyful, Pleased \\
& Proud, Surprised (positive), Triumphant \\
\midrule
Low Arousal, Positive Valence &
At Ease, Calm, Comfortable, Content \\
& Fulfilled, Grateful, Hopeful, Peaceful \\
& Relaxed, Relieved, Satisfied, Secure \\
& Serene, Sleepy, Tranquil, Well \\
\midrule
Low Arousal, Negative Valence &
Ashamed, Bored, Dejected, Depressed \\
& Disappointed, Dissatisfied, Droopy, Gloomy \\
& Guilty, Hopeless, Lonely, Miserable \\
& Sad, Tired, Worried \\
\midrule
High Arousal, Negative Valence &
Afraid, Agitated, Angry, Annoyed \\
& Anxious, Disgusted, Frustrated, Irritated \\
& Nervous, Overwhelmed, Panicked, Restless \\
& Shocked, Stressed, Tensed \\
\bottomrule
\end{tabular}
\caption{Emotions list as used in our application}
\label{tab:valence-arousal-emotions}
\end{table}


\end{document}